\documentclass[times,twocolumn,final]{elsarticle}

\usepackage{medima_preprint}
\usepackage{framed,multirow}
\usepackage{multirow}
\usepackage{amsmath,amssymb,amsfonts}
\usepackage{algorithmic}
\usepackage{graphicx}
\usepackage{enumitem}
\usepackage{textcomp}
\usepackage{subcaption}
\usepackage{caption}

\usepackage{latexsym}

\usepackage{url}
\usepackage{xcolor}
\usepackage{hyperref}

\definecolor{newcolor}{rgb}{.8,.349,.1}

\journal{Preprint}

\begin{document}

\verso{}

\begin{frontmatter}


\title{How does self-supervised pretraining improve robustness against noisy labels across various medical image classification datasets?}%


\author[1]{Bidur \snm{Khanal}\corref{cor1}}
\cortext[cor1]{Corresponding author is at Center for Imaging Science, RIT, USA }\ead{bk9618@rit.edu} 
\author[3]{Binod \snm{Bhattarai}}
\author[4]{Bishesh \snm{Khanal}}
\author[2,1]{Cristian \snm{Linte}}

\address[1]{Center for Imaging Science, RIT, Rochester, NY, USA}
\address[2]{Biomedical Engineering, RIT, Rochester, NY, USA}
\address[3]{University of Aberdeen, Aberdeen, UK}
\address[4]{NepAl Applied Mathematics and Informatics Institute for Research (NAAMII), Lalitpur, Nepal}


\received{}
\finalform{}
\accepted{}
\availableonline{}
\communicated{}

\begin{abstract}
Noisy labels can significantly impact medical image classification, particularly in deep learning, by corrupting learned features. Self-supervised pretraining, which doesn't rely on labeled data, can enhance robustness against noisy labels. However, this robustness varies based on factors like the number of classes, dataset complexity, and training size. In medical images, subtle inter-class differences and modality-specific characteristics add complexity. Previous research hasn't comprehensively explored the interplay between self-supervised learning and robustness against noisy labels in medical image classification, considering all these factors. In this study, we address three key questions: i) How does label noise impact various medical image classification datasets? ii) Which types of medical image datasets are more challenging to learn and more affected by label noise? iii) How do different self-supervised pretraining methods enhance robustness across various medical image datasets? Our results show that DermNet, among five datasets (Fetal plane, DermNet, COVID-DU-Ex, MURA, NCT-CRC-HE-100K), is the most challenging but exhibits greater robustness against noisy labels. Additionally, contrastive learning stands out among the eight self-supervised methods as the most effective approach to enhance robustness against noisy labels.
\end{abstract}


\end{frontmatter}

\section{Introduction}
\label{sec:introduction}

Accurately labeled data is pivotal to effectively train supervised deep learning methods for medical image classification. However, due to various factors such as outsourcing data labeling to non-experts for cost-effective annotation \citep{orting2020survey, radsch2023labelling}, automatic label generation from medical test reports \citep{irvin2019chexpert}, intentional label-flipping attacks \citep{xiao2012adversarial, steinhardt2017certified}, and even considerable variability among expert annotators \citep{radsch2023labelling}, datasets often exhibit high label noise. The presence of label noise during the training of deep learning models undermines their generalizability, resulting in suboptimal performance \citep{lee2019robust, zhang2021understanding, xia2021robust, khanal2021does, khanal2023investigating}. Consequently, researchers have directed their efforts towards developing robust methods capable of learning effectively even in the presence of noisy labels, a domain also referred to as learning with noisy labels (LNL). This includes studies in both natural image \citep{Patrini_2017_CVPR, hu2019simple, chen2019understanding, ren2018learning, liu2015classification, han2018co, Wei_2020_CVPR, song2019does, Li2020DivideMix:, wang2022promix} and medical image classification problems \citep{karimi2020deep, ju2022improving, xue2022robust, liu2021co, zhou2023combating, khanal2023improving}.

Many prior LNL methods directly rely on supervision from noisy labels to learn the feature extractors during the early epochs of training \citep{han2018co, Li2020DivideMix:, xia2021robust}. However, a significant drawback of such approach is the model's struggle to learn robust features in the initial phase—a challenge termed the warm-up obstacle \citep{zheltonozhskii2022contrast}. Recent solutions address this challenge by jointly using self-supervised learning, which doesn't rely on labels, and supervised learning with noisy labels \citep{li2022selective, ju2022improving, xue2022robust}. The self-supervised component ensures that the model doesn't drift towards learning corrupted features arising due to the supervision from noisy labels. However, a hurdle with such joint training is the addition of extra complexities in balancing various loss components and learning rates. Meanwhile, \cite{zheltonozhskii2022contrast, xue2022investigating} have shown that self-supervised pretraining alone can be equally effective in improving robustness against noisy labels. The advantage of such an approach is that the models can be independently pretrained with optimal settings and used directly with existing LNL methods, without dealing with the complexities of joint training. We demonstrated the effectiveness of this straightforward approach with medical image classification \citep{khanal2023improving}.

However, given the vast spectrum of self-supervised techniques covering contrastive learning, pretext-task-based approaches to generative methods, the choice of the best method of pretraining for improving robustness against noisy labels remains unclear. Additionally, robustness against noisy labels depends on various factors, including the number of classes, training dataset size, and dataset difficulty. Characteristics unique to medical images, such as subtle inter-class variability and modality-specific features, further contribute to this complexity.

In this work, we focused on understanding the impact of noisy labels in medical image classification and how self-supervised pretraining enhances robustness against it.  We began by revisiting the theoretical proof of the robustness against noisy labels, comparing it with empirical results that did not align, and highlighting the significance of learning robust features to fill the existing gap for improved robustness. Subsequently, we assessed five datasets to understand the ranking of datasets based on difficulty and their robustness against noisy labels. Finally, we conducted experiments with various self-supervised pretraining techniques to identify the optimal strategy for enhancing robustness against noisy labels. This is the complete extension of our previously published idea \citep{khanal2023improving}, encompassing in-depth studies, additional datasets, methods, and experimental analysis. Our contributions can be summarized as follows:

\begin{itemize}
    \item Studying five different medical image classification 2D datasets, encompassing X-ray, ultrasound, and RGB images, to comprehend how factors like dataset size, the number of classes, and dataset difficulty influence robustness against noisy labels.

\item Examining eight self-supervised learning techniques, including contrastive learning-based, pretext task-based, and generative approaches, to determine the most effective method for pretraining to enhance robustness against noisy labels across the five datasets.

\item Discussing the best strategy and scenarios where self-supervised pretraining proves beneficial, supported by additional ablation studies.
\end{itemize}

\section{Related Works on Learning with Noisy Labels in Medical Image Classification}

Numerous techniques have been proposed to enhance the resilience of medical image classifiers in the presence of noisy labels. These methods encompass a variety of strategies, including the utilization of label smoothing \citep{pham2021interpreting} for conditions like thoracic diseases, pancreatic and skin cancers, breast tumors, and retinal diseases, as well as architectural modifications like the incorporation of a noise layer \citep{dgani2018training}. Additionally, some approaches involve sample re-weighting techniques \citep{le2019pancreatic,xue2019robust}, uncertainty-based methodologies \citep{ju2022improving}, and various mathematical techniques such as PCA, low-rank representation, graph regularization, among others \citep{ying2023covid}.

Furthermore, methods like consistency regularization and disentangled distribution learning \citep{zhou2023combating}, student-teacher co-training \citep{xue2022robust}, and co-correcting in conjunction with curriculum learning \citep{liu2021co} have been explored. Nevertheless, it is worth noting that, despite incorporating elements from self-supervised learning, no research has yet investigated the potential impact of exclusively employing self-supervised pretraining as a means to enhance robustness against noisy labels.

\cite{hendrycks2019using} have shown that pretraining can improve robustness to noisy labels. \cite{zheltonozhskii2022contrast,xue2022investigating} demonstrated that training with self-supervised learning can further enhance existing LNL methods. We then showed that self-supervised pretraining can also improve medical image classification in the presence of noisy labels \citep{khanal2023improving}. However, our previous work lacked investigation in a wide range of medical image datasets and self-supervised techniques. Additionally, no prior works provide insight into understanding how dataset difficulty and the number of classes influence the robustness against noisy labels in medical image classification and under what conditions self-supervised pretraining proves beneficial.

\section{Method}
The method is divided into subsections: i) Problem setup and ii) Proposed Pipeline.  In the problem setup section, we mathematically introduce label noise, present theoretical assumptions, and investigate the impact of noisy labels across various datasets by comparing datasets based on difficulty and robustness against noisy labels. In the proposed pipeline section, we describe our approach to improving robustness against noisy labels using self-supervised pretraining and existing LNL methods, where we thoroughly investigate various self-supervised methods.

\subsection{Problem Setup}
In the problem setup section, we first discus the how label noise is injected into a dataset. Then, we discuss the impact of noisy labels both theroritically and emprically thorugh our experiments to point out the difference in theoritical assumption and empirical results. And then establish the motivation for using self-supervised pretraining. After that we empirically assess and try to rank datasets based on difficulty through various metrices and then test the robustness of each datasets against respective noisy labels to understand if there exists a relationship between dataset difficulty and robustness to noisy labels.
\subsubsection{Label Noise Injection}
\label{label noise injection}
Consider a clean training dataset $D=\left\{\left(\boldsymbol{x}_i, y_i^{*}\right)\right\}_{i=1}^n$ with $n$ samples, where each $\left(\boldsymbol{x}_i, y_i^{*}\right) \in(X \times Y^{*})$. Here, $X \in \mathbb{R}^{d \times d }$ represent the input images, and $Y^* \in \{1,2, \ldots, c\}$ represent the corresponding true labels, where c is the number of classes. We inject label noise into the existing dataset by randomly flipping labels to incorrect labels with certain probability. There are two approaches to injecting label noise: 1) the likelihood of choosing each incorrect class is equal, also referred to as symmetrical label noise; 2) some incorrect classes have a higher likelihood of being chosen for a given true class, also referred to as asymmetrical or class-dependent label noise. When injecting symmetric label noise into a sample $(x_i, y_i^*)$, its label $y_i^*$ is changed to $y_i \overset{\epsilon}{\sim} Y^* \setminus y_{i}^*$ with a probability $\epsilon$, also referred as label noise rate. Here, $Y^* \setminus y_{i}^*$ represents any class label within the closed-set $Y^*$ other than the true label. But for class-dependent label noise, $y_i^*$ is changed to $y_{i} \overset{\epsilon}{\sim} S$, with $\epsilon$. Here, $S \in \{1,2,\ldots, s\}$, where $S$ is subset of classes that are likely to be confused with $y_i^*$ (also referred to as dependent classes). The number of classes in $S$ is referred to as \textit{spread} (s). In our experiments, we inject label noise at various rates, specifically $\epsilon \in \{0.5, 0.6, 0.7, 0.8\}$ for symmetrical case, and $\epsilon \in \{0.3,0.4,0.5, 0.6, 0.7\}$ for class-dependent case, to simulate scenarios with high noise rates.

\begin{figure*}[h!]
\centering
    \includegraphics[width=1\linewidth]{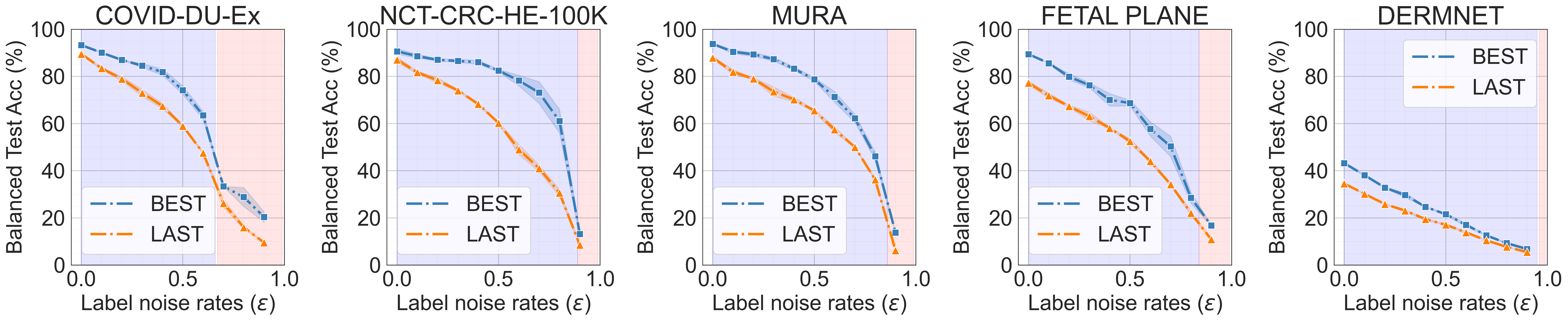}
    \label{fig:auroc}
    \vspace{-1.5em}
\caption{The assessment of test performance relative to training label noise rate (noise probability). BEST represents peak performance, and LAST denotes the average over the last five epochs. The blue region indicates noise rates below the flipping threshold, while the red region signifies rates above the threshold.} 
\label{fig:noise_impact}
\end{figure*}

\subsubsection{Impact of Noisy Labels}
\label{impact-of-noisy-labels}

Let's consider a simple medical image classification task: categorizing chest X-ray images into specific pathological conditions like pneumonia, COVID-19 infection, and more. Assume a clean dataset $D=\left\{\left(\boldsymbol{x}_i, y_i^*\right)\right\}_{i=1}^n$ containing $n$ training samples, such that $\left(\boldsymbol{x}_i, y_i^*\right) \in(X \times Y^*)$, where $X \in \mathbb{R}^{d \times d }$ represent input images, $Y^* \in \{1,2, \ldots, c\}$ represent possible pathological conditions with $c$ representing the total number of conditions or classes.

Using a Bayesian interpretation, if we know the prior probability $P[Y^* = k] = \pi_k$ for each pathological condition $k \in \{1,2, \ldots, c\}$ and the likelihood $P[X = x|Y^* = k]$ of observing a chest X-ray image $X = x$ given condition $k$, we can calculate the posterior probability $P[Y^* = k \mid X=x]$. 

Now, let's assume we observe noisy labels $Y$ instead of true labels $Y^*$. The conditional probability for noisy labels, $P[Y=i \mid Y^* = k, X=x] = \eta_{ki}(x)$ where $\sum_{i=1}^c \eta_{ki}(x)=1$, represents the probability of observing noisy label $i$ given true label $k$. In this scenario, the noisy posterior $P[Y = k \mid X=x]$ is the probability of observing label $Y = k$ given X-ray image $X=x$. The distribution of noisy labels can be either symmetrical for all classes or class-dependent.

\begin{itemize}[leftmargin=*]
    
\item \textbf{Symmetrical label noise:}
With label noise in the range $0 \leq \epsilon \leq 1$ (probability of flipping true labels), the noisy label distribution is given by:
\begin{equation}
\label{noisy_label_dist}
\eta_{k i}= \begin{cases}1-\epsilon, & \text { if } i=k \\ \frac{\epsilon}{c-1}, & \text { if } i \neq k\end{cases}
\end{equation}
\cite{oyen2022robustness} derived that the clean test accuracy drops to half at $\epsilon = \frac{c-1}{c}$ (referred to as the \textit{flipping threshold}), implying that even at a high noise rate (e.g., $\epsilon = 0.88$ for $c = 9$), the classifier remains robust. However, our empirical results (Fig. \ref{fig:noise_impact}) do not fully support this claim, as performance deteriorates before reaching the flipping threshold.


We hypothesize that this issue arises from suboptimal feature learning. In deep learning, there is a dependency $X \Longrightarrow Z \Longrightarrow Y$, where $Z \in \mathbb{R}^D$ represents the feature representation obtained as $Z = G_\theta(X)$, depending on model parameters $\theta$. When $G_\theta$ becomes corrupted during training with noisy labels, the feature representation becomes subpar, and the theoretically determined posterior $P\left[Y=k \mid X=x\right]$ cannot be achieved. Thus, achieving an optimal $G_\theta$ is crucial for ensuring theoretical robustness. However, supervised learning with noisy labels often degrades the feature representation \citep{zheltonozhskii2022contrast, li2022selective}. Therefore, self-supervised learning can serve as a solution to address the problem of suboptimal feature learning and, consequently, enhance the classifier's robustness against noisy labels.

\item \textbf{Class-dependent label noise:}\label{class-dependent-theory} In the context of class-dependent label noise, a unique scenario arises where the probability of a label flipping from a true class $k$ to a non-dependent class $i$ is typically zero ($\eta_{k} = 0$). However, the likelihood of the label being flipped to other dependent classes remains non-zero ($\eta_{ki} \neq 0$). For instance, an expert radiologist might mislabel a chest X-ray image of a pneumonia patient as COVID-19 but is less likely to label it as normal. The number of dependent classes susceptible to mislabeling for the true class $k$ is referred to as \textit{spread}($s$) \citep{oyen2022robustness}. In the case of class-dependent label noise, the distribution of noisy labels is as follows:

\begin{equation}
\eta_{k i}= \begin{cases}1-\epsilon, & \text { for } i=k \\ \epsilon \cdot t_{k i} / s, & \forall i \in\{1, \ldots, c\} \backslash\{k\} \end{cases}
\end{equation}
\begin{equation*}
\\\text { where } \sum_{i \neq k}\left|t_{k i}\right|^0=s \text { and } \sum_{i \neq k} t_{k i}=1    
\end{equation*}

\cite{oyen2022robustness} theoretically showed that the label flipping noise threshold for class-dependent label noise can be as low as $\epsilon = 1/2$, depending on the \textit{spread} ($s$). This means that even with an optimal mapping function $Z = g(X;\theta)$, the classifier's performance for $c = 9$ classes could significantly deteriorate at $\epsilon = 0.5$. Therefore, theoretical robustness is low for class-dependent label noise, which was confirmed by our experimental results later in Section \ref{class-dependent label noise}.

\end{itemize}
\subsubsection{Assess Dataset Difficulty:}
\label{dataset difficulty}The impact of noisy labels on a dataset's classification model depends on the dataset's inherent characteristics. Some datasets display clear class separations, making them easier to train, while others lack distinct decision boundaries, particularly evident in medical image classification with subtle inter-class pixel-level variations. Quantifying and comparing dataset difficulty can be challenging, given varying factors like class count and dataset size, directly influencing the learning process. For example, some datasets train classification models well due to their large training samples despite having difficult targets, while others do not optimize training because they lack sufficient training samples despite being easier. Furthermore, if the test distribution significantly deviates from the training set, even a well-trained model may perform poorly during evaluation. These elements collectively shape classifier training. In this paper, we aim to impartially quantify dataset difficulty by reducing the influence of dataset size and class count, addressing the question: \textbf{\textit{Which dataset is inherently difficult?}}. To answer, we analyze test performance with varying training dataset sizes, maintaining constant class groupings. We also assess test performance with different class groupings while keeping the dataset size consistent. Furthermore, we evaluate class separability scores for both training and test sets.

\begin{itemize}[leftmargin=*]
\item \textbf{Test performance vs. class numbers:}
\label{vsclass}
We maintained a consistent dataset size for all datasets, regardless of the number of classes, by randomly sampling 7,000 training samples. Classes were grouped into three, six, seven, or thirteen, as allowed by the dataset, in both training and test sets. Additional details on class grouping are in Experimental section \ref{class-grouping}. For instance, COVID-QU-Ex, with only three classes, couldn't be further grouped, while DermNet, with 23 classes, allowed grouping into three, six, seven, and thirteen. We trained ResNet18 with standard cross-entropy loss using the same training hyperparameters as detailed in Experimental section \ref{learning with noisy labels hyperparameters}. We conducted 6-fold cross-validation on different 7000-sample sets for each dataset, evaluating on the respective grouped-test set. Referencing Fig. \ref{fig:testvsclass}, which illustrates test F1-scores across different class numbers, it becomes evident that DermNet exhibited the poorest performance, while MURA consistently outperformed the rest across various class counts. This underscores the higher difficulty level associated with DermNet compared to the other datasets on this metric.


\begin{figure}[h!]
\centering
\includegraphics[width=0.8\linewidth]{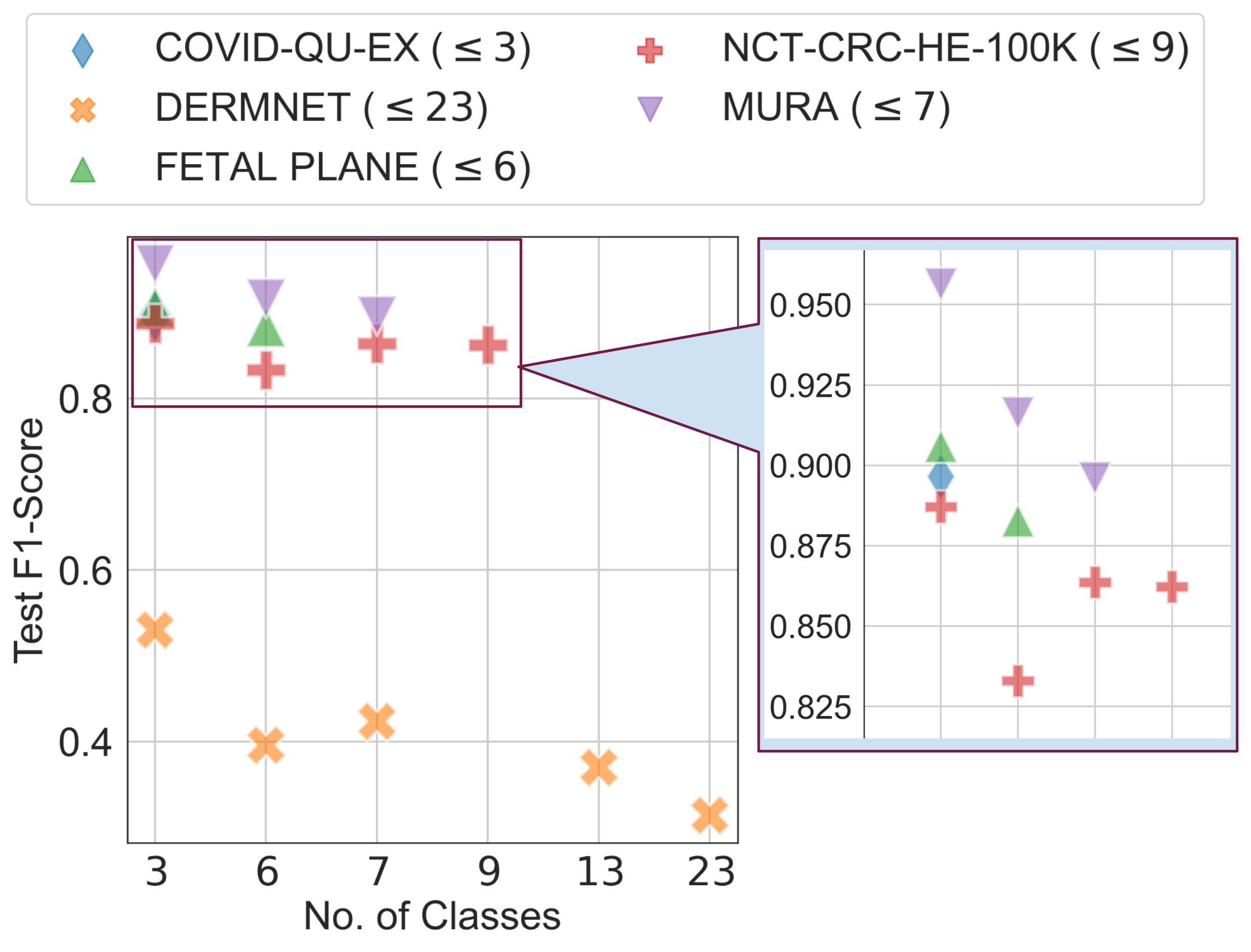}
    \caption{Comparison of test performance of models trained with varying numbers of classes while keeping the same 7000 training samples across five datasets. The classes are varied by grouping them to reduce the original class count. The symbol $\leq$ denotes the actual number of classes in the dataset, which cannot be exceeded.}
\label{fig:testvsclass}
\end{figure}
  
\item \textbf{Test performance vs. dataset size:}
\label{vsdataset}
In this setup, we standardized the number of classes for each dataset to three using the class grouping method (Experimental section \ref{class-grouping}). We then trained each dataset at different sizes, including ${1000, 3000, 5000, 7000, 15000, 27000, 36000, 100000}$, up to the original training size limit. For example, the Fetal dataset, with 7129 training samples, couldn't exceed a size of 7000, while DermNet could be trained with sizes above 15,000. Only NCT-CRC-HE-100K allows training at all specified sizes. We followed the same architecture and training configuration as in preceding section. Similarly, we conducted 6-fold cross-validation, evaluating each fold on the grouped-test set. According to Fig. \ref{testvssize}, DermNet exhibits the lowest performance, while MURA demonstrates the highest performance, confirming the conclusion of preceding section.

\begin{figure}[h!]
\centering
\includegraphics[width=0.8\linewidth]{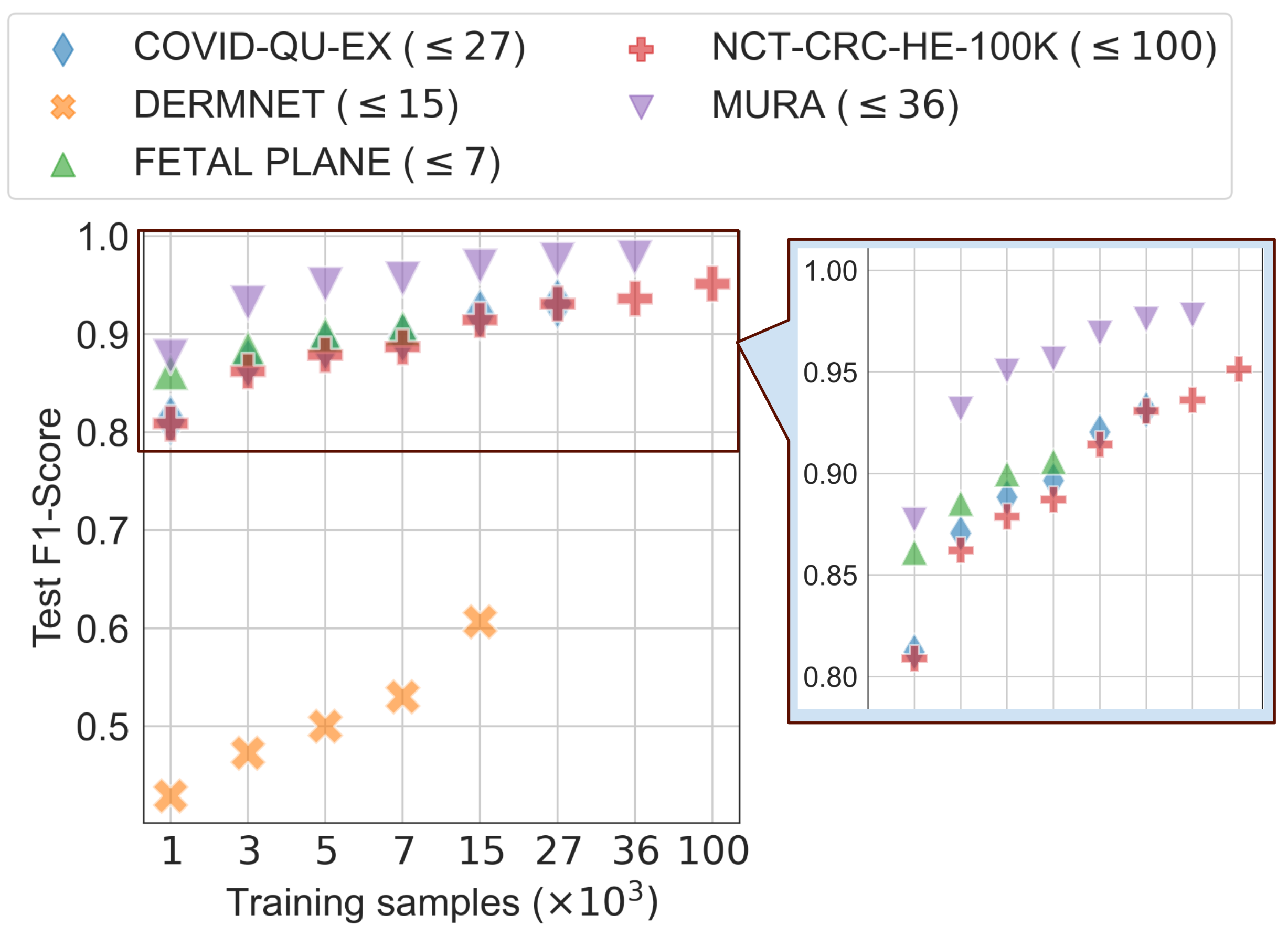}
    \caption{Comparison of test performance of models trained with varying numbers of training samples while keeping the same number of classes (three). The symbol $\leq$ denotes the number of training samples (in $10^{3}$) present in the dataset, which cannot be exceeded.}
\label{testvssize}
\end{figure}



\item \textbf{Fisher's Class Separability Score (CSS):}
\label{fisher CSS}We assessed class separability in each dataset using Fisher's Class Separability Score (CSS) \citep{bishop2006pattern}, which considers both sample mean and variance and normalizes the score to account for the class count. CSS is computed from the feature embeddings of a trained model. To ensure a fair comparison of CSS across datasets, we map all datasets to a common feature embedding. Additionally, instead of directly relying on feature embeddings from a model trained on an out-of-domain dataset like ImageNet, we jointly fine-tuned the ImageNet pretrained model on all five datasets, as shown in Section \ref{joint feature embedding}. Fig. \ref{fig:fisherscore} shows the Fisher's CSS for all datasets. DermNet consistently exhibits the lowest Fisher's CSS. Notably, CSS varies between test and training sets for other datasets, indicating that class separability in the training set may not necessarily extend to the test set.

\begin{figure}[h!]
\centering
\includegraphics[width=0.6\linewidth]{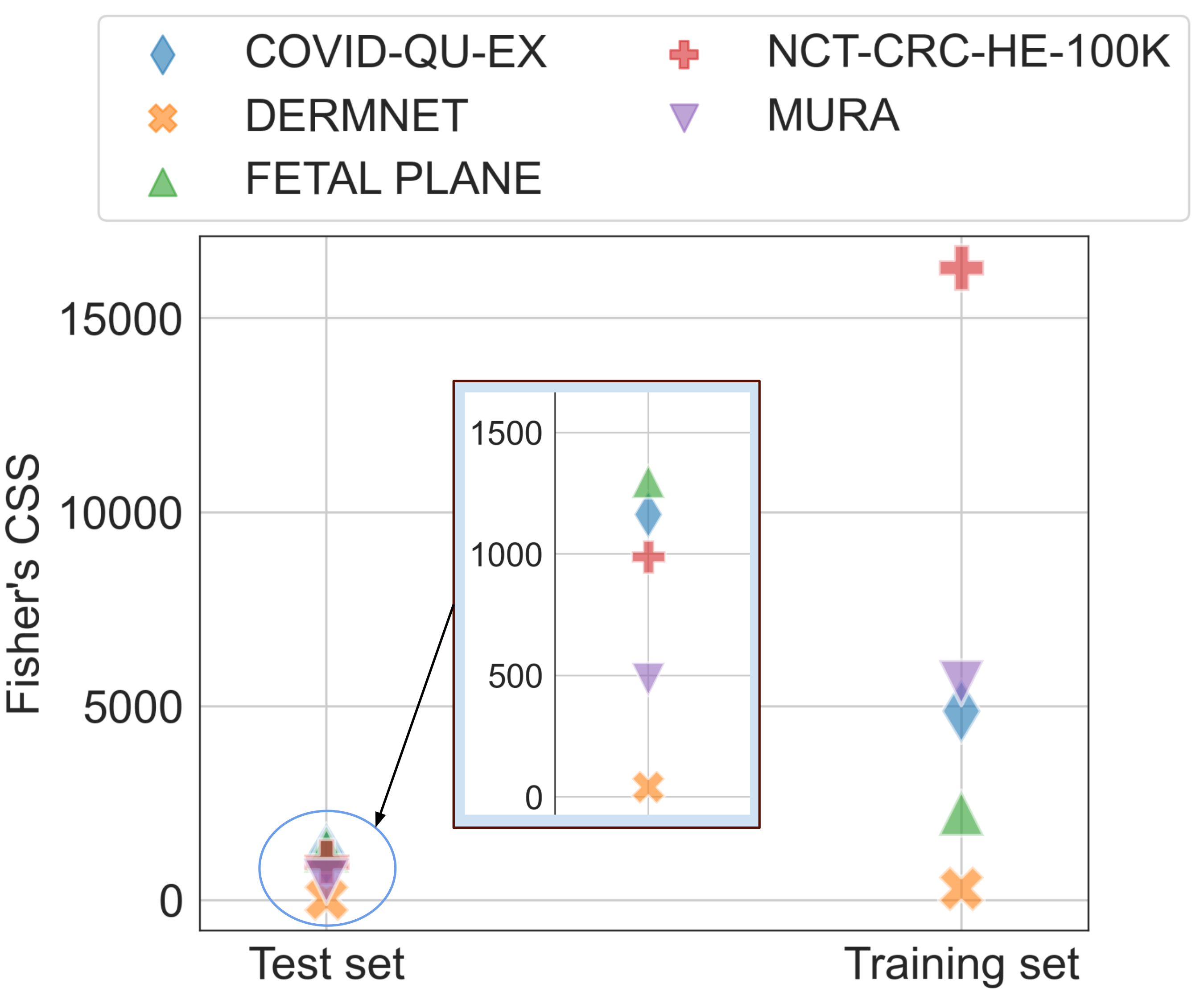}
\caption{Comparison of Fisher's Class Separability Score (CSS) across all five datasets in both the training set and test set.}
\label{fig:fisherscore}
\end{figure}


\end{itemize}
\subsubsection{Robustness to Label Noise}
\label{robustnesstolabelnoise}

We evaluated dataset robustness against noisy labels using a metric that measures the relative drop in test performance (TP) across different label noise rates. We chose the F1-score to evaluate TP for this analysis. The robustness score (R) is calculated as $R = \frac{\sum_{\eta=0}^{H} |1|}{\sum_{\eta=0}^{H} (\text{TP}_{0} - \text{TP}_{\eta})}$.
\begin{figure}[h!]
\centering
\includegraphics[width=0.6\linewidth]{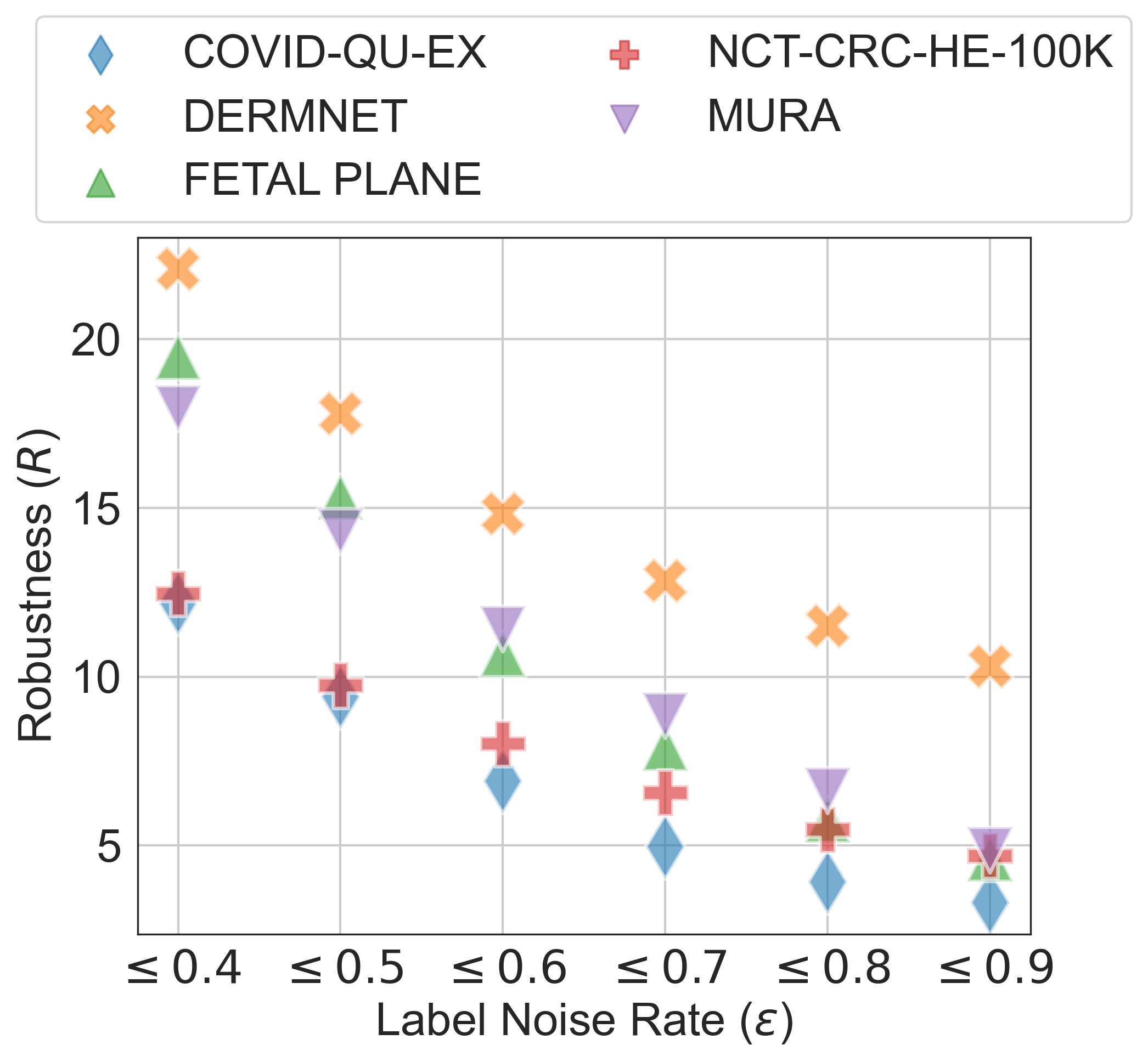}
\caption{Comparison of the robustness score across various datasets. The higher value of R denotes relatively greater robustness against noisy labels. R for $\epsilon \leq 0.4$ decipts that the robustness score was computed for label noise range {0,0.4}.}
\label{fig:robustness_metric}
\end{figure}

For a fair comparison, we maintained uniform training dataset sizes and class counts across all datasets. Grouping classes into three as detailed in Section \ref{vsclass}, we randomly sampled 7000 training samples for ResNet18 training across all noise rates. We computed 6-fold cross-validation on different 7000-sample sets for each dataset. Fig. \ref{fig:robustness_metric} presents the robustness score, with higher values indicating greater resilience to label noise. DermNet displays the least sensitivity to label noise, while COVID-QU-Ex is the most affected. It should be noted that DermNet, being inherently challenging, shows lower performance even without label noise, resulting in less relative performance degradation due to label noise compared to other datasets.


\subsection{Proposed Pipeline}
\label{method_section}
\begin{figure}[h!]
\centering
\includegraphics[width=1\linewidth]{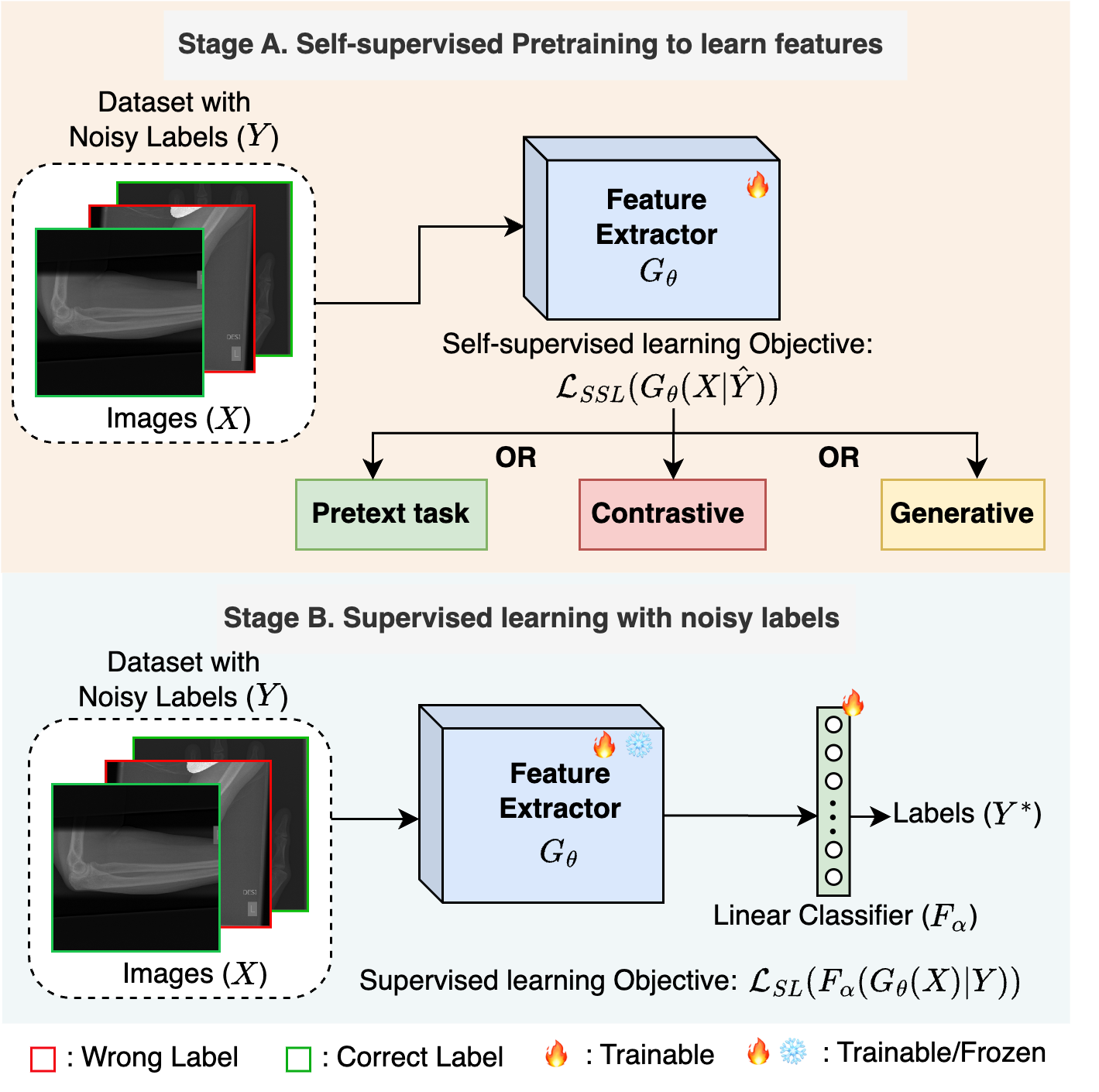}
    \caption{The overall pipeline consists of two stages: a) self-supervised pretraining to learn a robust feature extractor \(G_\theta\), and b) supervised training on noisy labels to build a robust classifier (\(G_\theta\); \(F_\alpha\)). During self-supervised learning, there is no use of the provided labels \(Y\); instead, it relies on self-generated pseudo labels \(\hat{Y}\). We explore various self-supervised learning objectives based on pretext tasks, contrastive learning, and generative methods. Supervised training employs the LNL method, which robustly trains the classifier using the noisy labels \(Y\).}
\label{fig:method}
\end{figure}

Our goal is to train a robust classifier $F_\alpha(G_\theta(X)|Y)$, where instead of observing all true labels Y*, we observe noisy labels Y. We employ a two-stage pipeline: i) Self-supervised pretraining phase, and ii) Supervised training using LNL methods. In the first stage, we pretrain the model using self-supervised learning approach on the given dataset. In the second stage, we adapt the pretrained backbone model to train a supervised classifier on the downstream medical image classification dataset containing noisy labels. Fig. \ref{fig:method} shows the overall pipeline of our proposed approach.

\subsubsection{Self-supervised Pretraining}

Self-supervised pretraining does not rely on provided ground truth labels; instead, it utilizes alternative forms of self-supervision, employing pseudo labels. Mathematically, the objective of self-supervised pretraining can be formulated as:

\begin{equation}
    \theta^* = \underset{\theta}{\operatorname{argmin}}\sum_{(X_i, Y_i) \in \mathcal{D}} \mathcal{L}_{SSL}(G_\theta(X_i), \hat{Y}_i)
\end{equation}
where \(X_i\) represents the input, \(Y_i\) stands for the provided label, \(\hat{Y}_i\) denotes the self-generated pseudo label, \(G_\theta\) represents the model parameterized by $\theta$, and \(\mathcal{L}_{SSL}\) represents the self-supervised loss function.

There are several self-supervised techniques, ranging from solving simple pretext tasks \citep{zhang2016colorful,gidaris2018unsupervised,doersch2015unsupervised,noroozi2016unsupervised}, to contrastive learning \citep{chen2020simple,he2020momentum,zbontar2021barlow,caron2020unsupervised,bardes2021vicreg}, to generative approaches \citep{kingma2013auto,goodfellow2014generative,pathak2016context,donahue2019large}. We investigated several techniques from these three categories to investigate which self-supervised technique offers greater robustness against noisy labels.

\begin{itemize}[leftmargin=*]
\item \textbf{Pretext tasks:}
\label{self-supervision_pretext_task}
Pretext task-based self-supervised learning employs auxiliary tasks to learn useful representations or features from the data without relying on provided labels. The objective function is formulated as:

\begin{equation}
    \hat{\alpha}^*; \theta^{*}_{\text{pretext}} = \underset{\hat{\alpha}; \theta}{\operatorname{argmin}} \sum_{X_i \in \mathcal{D}} \mathcal{L}(F_{\hat{\alpha}}(G_\theta(X_i)), \hat{Y_i})
\end{equation}

where \(\hat{Y}_i\) represents pseudo label. For example, in a rotation-based task, the objective is to predict the orientation of a self-rotated input image, and the pseudo labels are the degrees of rotation (\(0^\circ, 90^\circ, 180^\circ, 270^\circ\)). Similarly, other tasks use different forms of pseudo-labels. Once the model is trained, our primary interest lies in \(G_\theta\), and \(F_{\hat{\alpha}}\) is discarded.

In this study, we selected three pretext tasks: \textit{Rotation Prediction} \citep{gidaris2018unsupervised}, \textit{Jigsaw Puzzle} \citep{noroozi2016unsupervised}, and \textit{Jigmag Puzzle} \citep{koohbanani2021self}. While \textit{Rotation Prediction} and \textit{Jigsaw Puzzle} are commonly used methods in the literature, \textit{Jigmag Puzzle} was introduced to address the challenges posed by medical images that lack consistency in shape. For \textit{Jigsaw Puzzle}, an input image is divided into a square grid of patches, and these patches are randomly shuffled. A model is then trained to predict the arrangement of the shuffled patches in the grid. \textit{Jigmag Puzzle}, originally designed for histopathology images, involves magnifying the image at different random locations with varying magnifications and arranging them in a grid. Then, a model is trained to predict the arrangement of these magnified patches.


\item \textbf{Contrastive learning: }
The contrastive learning objective trains a model to distinguish between positive pairs (augmented views of the same sample) and negative pairs (other samples) in the data. The objective is to bring positive pairs closer together in the feature space while pushing negative pairs further apart. The objective function of simple contrastive learning is as follows:

\begin{equation}
    \mathcal{L}_{\text{contrastive}}(i, j) = -\log\frac{\exp(sim(Z_i, Z_j) / \tau)}{\sum_{k=1, k \neq i}^{2N} \exp(sim(Z_i, Z_k) / \tau)}
\end{equation}

Here, $sim$ measures the similarity between the feature representations \(Z_i\) and \(Z_j\), which are obtained from \(Z = G_\theta(X)\). \(\tau\) is an adjustable temperature parameter, and \(N\) represents the total number of samples in a batch.

Out of various approaches, we have focused our study on three state-of-the-art contrastive methods: \textit{SimCLR} \citep{chen2020simple}, \textit{MoCo} \citep{he2020momentum}, and \textit{Barlow Twins} \citep{zbontar2021barlow}. \textit{SimCLR} employs a contrastive loss that maximizes the similarity between the augmentated views of the same image (positive pairs), while minimizing the similarity with other images (negative pairs) within the mini-batch. The similarity is measured in low-dimensional embedding space with cosine distance. The performance of SimCLR is limited by the number of negatives that be accommodated in a mini-batch, which is dictated by GPU memory. \textit{MoCo} was proposed to overcome this challenge by introducing a memory bank for storing representation of negative pairs. It utilizes a lookup dictionary, constructed as a queue, to save a large number of encoded examples as keys. In \textit{MoCo}, an encoded sample, referred to as a query, is compared against the keys using a contrastive loss similar to SimCLR. It employs two separate encoders for keys and queries, and updates the key encoder using the momentum of the query encoder. On the other hand, \textit{Barlow Twins} is a negative-free approach that doesnot require negative pairs. It aims to minimize the redundancy between the embedding vectors of two augmented views of the same image. To achieve this, it computes the cross-correlation matrix from the embedding vectors and enforces an objective function that equates the diagonal terms of the matrix to 1, while minimizing the off-diagonal terms towards 0.


\item \textbf{Generative approaches:}
The generative approach, often referred to as an unsupervised learning technique, is trained with pixel-level image reconstruction or generation objectives and doesn't require ground truth labels, nor does it use pseudo labels. Its formulation differs from discriminative approaches. It employs an encoder and decoder architecture, where the encoder encodes the input image, and the decoder reconstructs it. The simplest form of the generative model can be represented as:

\begin{equation}
    \alpha^*; \theta^{*} = \underset{\alpha; \theta}{\operatorname{argmin}} \sum_{X_i \in \mathcal{D}} \mathcal{L}_{\text{generative}}(D_{\alpha}(G_\theta(X_i)), X_i)
\end{equation}

Here, \(G_\theta\) is the encoder, and \(D_\alpha\) is the decoder. We retain only \(G_\theta\) for downstream supervised classification tasks and discard \(D_\alpha\) after training.

Generative approaches cover simple autoencoders \citep{bishop2006pattern} to VAEs \citep{kingma2013auto} to GAN-based approaches \citep{goodfellow2014generative,pathak2016context,donahue2019large}. In this study, we focus on two generative approaches: the vanilla VAE \citep{kingma2013auto} and a GAN-based approach called BigBiGAN \citep{donahue2019large}. The VAE encodes the input image as a distribution over the latent space. To generate a new image, a point from the latent distribution is sampled and decoded. It is trained with a reconstruction loss and a KLD regularizer that enforces the latent distribution towards a normal distribution. BigBiGAN, a successor of BiGAN, was proposed with the objective to enhance representational learning capability. Unlike typical GANs that use a generator to generate images from a latent vector, BigBiGAN also has an encoder that encodes the input image to a latent space. A discriminator is adversarially trained to discriminate between the generator and latent distribution pair versus the encoder and input distribution pair. We are only interested in the encoder part.
\end{itemize}

\subsubsection{Supervised Learning with LNL}

In this phase, we employ supervised learning to train our model for medical image classification using noisy labels. The supervised training utilizes the feature extractor pretrained with self-supervised learning in Phase 1. The objective function of this phase is formulated as:

\begin{equation}
    \alpha^*; \theta^* = \underset{\alpha; \theta}{\operatorname{argmin}} \sum_{(X_i, Y_i) \in \mathcal{D}} \mathcal{L}_{SL}(F_\alpha(G_\theta(X_i)), Y_i)
\end{equation}

where \(X_i\) represents the input, \(Y_i\) stands for the provided label, \(G_\theta\) represents the pretrained feature extractor parameterized by \(\theta\), \(F_{\alpha}\) is a new linear classifier parametrized by \(\alpha\), and \(\mathcal{L}_{SL}\) represents the supervised loss function.

Rather than employing a standard cross-entropy loss for classifier training, we opt for existing LNL methods, known for their robustness. Specifically, we have chosen two state-of-the-art LNL methods: Co-teaching (CT) \citep{han2018co} and Dividemix (DM) \citep{Li2020DivideMix:}. Co-teaching (CT) utilizes loss-based sample selection, ranking samples by their learning loss, and exclusively trains the model with the top $\tau$ samples exhibiting the least loss values. In contrast, Dividemix (DM), in addition to loss-based selection, rectifies noisy samples by treating them as unlabeled data and, after rectification, includes them in the training process.  We chose these methods because both rely on an initial warm-up phase, where an effective feature extractor is crucial for successful loss-based selection.

\section{Datasets}

\begin{itemize}[leftmargin=*]
    \item \textbf{NCT-CRC-HE-100K:} NCT-CRC-HE-100K is a histopathological image dataset that contains $224 \times 224$ patches extracted from stained tissue slides, both cancerous and normal, totaling 100,000 RGB images \citep{kather2019predicting,kather_jakob_nikolas_2018_1214456}. This dataset encompasses nine distinct categories, including adipose tissue, lymphocytes, mucus, and others. Notably, the test set employs a separate CRC-VAL-HE-7K dataset, containing 7,180 images.
\item \textbf{MURA:} MURA is a large collection of musculoskeletal radiographs, containing a total of 40,561 images of varying sizes obtained from 14,863 studies, which involved 12,173 patients \citep{rajpurkar2017mura}. Each image is categorized into one of seven classes, such as shoulder, humerus, elbow, and others. The original dataset has already been divided into training and validation sets, consisting of 36,808 images from 11,184 patients for training and 3,197 images from 783 patients for validation, ensuring that there is no patient overlap. In our study, we utilized the validation set as our test set.
\item \textbf{COVID-QU-Ex:} COVID-DU-Ex comprises chest X-ray images sourced from several patients \citep{anas_2022}. The dataset includes 27,132 training images, of which 9,561 as COVID-19 are classified as COVID-19 instances, 9,010 as Non-COVID-19, and 8,561 classified as Normal. Furthermore, 6,788 samples are provided separately as test set.
\item \textbf{DermNet:} DermNet dataset contains a total of 19,559 dermatology RGB images of varying size splitted into 15557 and 4002 training and test set respectively \footnote{https://www.kaggle.com/datasets/shubhamgoel27/dermnet}. Each image belong to one of 23 skin conditions such as  Acne, Melanoma, Poison Ivy, Psoriasis, Eczema, etc.
\item \textbf{Maternal-fetal US:} Maternal-fetal US dataset includes ultrasound images from maternal-fetal scans, each manually labeled by expert clinicians \citep{burgos2020evaluation}. The dataset consists of 12,400 ultrasound images from 1,792 patients, split into patient-wise non-overlapping sets of 7,129 training images and 5,271 test images. These images are classified into six classes: Fetal Abdomen, Fetal Brain, Fetal Femur, Fetal Thorax, Mother's Cervix, and a general class labeled as 'other.'
\end{itemize}

\section{Experiments}
\subsection{Implementation Details for Problem Setup}

In this section, we delve into the implementation details of aspects covered in the Problem Setup section: joint feature space creation, class grouping, and label noise injection.
\begin{table*}[!h]
\scriptsize
\setlength{\tabcolsep}{4pt}
\centering
\caption{Grouping of classes for Fetal dataset}
\label{Fetal-class-grouping}
\begin{tabular}{l}
\hline 3 classes \\\hline
Group 1: [``Fetal abdomen", ``Fetal brain", ``Fetal femur", ``Fetal thorax"] \\
Group 2: [``Maternal cervix"] \\
Group 3: [``Other"] \\
\hline
\end{tabular}

\end{table*}
\begin{table*}[!h]
\tiny
\centering
\scriptsize
\setlength{\tabcolsep}{4pt}
\caption{Grouping of classes for NCT-CRC-HE-100K dataset}
\label{NCT-CRC-HE-100K-class-grouping}
\begin{tabular}{lll}
\hline 3 classes &                                             6 classes &                                             7 classes \\
\hline
Group 1: [``Adipose", ``Smooth muscle", ``Colon mucosa"] & Group 1: [``Adipose", ``Smooth muscle", ``Colon mucosa"] & Group 1: [``Adipose", ``Smooth muscle", ``Colon mucosa"] \\
Group 2: [``Background", ``Debris", ``Lymphocytes", ``Mucus"] &  Group 2: [``Background", ``Mucus"] &   Group 2: [``Background"] \\
Group 3: [``Cancer stroma", ``Adenocarcinoma"] & Group 3: [``Debris"] & Group 3: [``Debris"] \\
&  Group 4: [``Lymphocytes"] &  Group 4: [``Lymphocytes"] \\ &                            Group 5: [``Cancer stroma"] &  Group 5: [``Mucus"] \\
&  Group 6: [``Adenocarcinoma"] &  Group 6: [``Cancer stroma"] \\
&  &    Group 7: [``Adenocarcinoma"] \\
\hline
\end{tabular}
\end{table*}

\begin{table*}[!h]
\tiny
\scriptsize
\centering
\setlength{\tabcolsep}{4pt}
\caption{Grouping of classes for Mura dataset}
\label{Mura-class-grouping}
\begin{tabular}{ll}
\hline
3 classes &        6 classes \\\hline
Group 1: [``XR\_SHOULDER", ``XR\_HUMERUS"] &   Group 1: [``XR\_SHOULDER"] \\
Group 2: [``XR\_FINGER", ``XR\_HAND"] & Group 2: [``XR\_HUMERUS", ``XR\_FOREARM"] \\
Group 3: [``XR\_WRIST", ``XR\_FOREARM", ``XR\_ELBOW"] &  Group 3: [``XR\_FINGER"] \\
&  Group 4: [``XR\_WRIST"] \\
&  Group 5: [``XR\_HAND"] \\
&  Group 6: [``XR\_ELBOW"] \\
\hline
\end{tabular}

\end{table*}
\begin{table*}[!h]
\tiny
\scriptsize
\centering
\setlength{\tabcolsep}{4pt}
\caption{Grouping of classes for Dermnet dataset}
\label{Dermnet-class-grouping}
\begin{tabular}{l}
\hline    3 classes \\ \hline
Group 1: [``Acne", ``Atopic", ``Cellulitis", ``Eczema", ``Poison Ivy", ``Psoriasis", ``Seborrheic"] \\
Group 2: [``Herpes HPV", ``Scabies Lyme", ``Tinea Ringworm", ``Warts Molluscum"] \\
Group 3: [``Actinic", ``Bullous", ``Exanthems", ``Hair Loss", ``Light Diseases", ``Lupus", ``Melanoma", ``Nail Fungus", ``Systemic", ``Urticaria Hives", ``Vascular Tumors", ``Vasculitis"] \\ \hline 
6 classes \\ \hline
Group 1: [``Atopic", ``Eczema", ``Exanthems", ``Psoriasis", ``Vasculitis"] \\
Group 2: [``Cellulitis", ``Herpes HPV", ``Lupus", ``Scabies Lyme", ``Tinea Ringworm"] \\
Group 3: [``Actinic", ``Melanoma"] \\
Group 4: [``Bullous", ``Seborrheic", ``Vascular Tumors"] \\
Group 5: [``Poison Ivy", ``Urticaria Hives"] \\
Group 6: [``Acne", ``Hair Loss", ``Light Diseases", ``Nail Fungus", ``Systemic", ``Warts Molluscum"] \\ \hline
7 classes \\ \hline
Group 1: [``Atopic", ``Eczema", ``Exanthems", ``Psoriasis", ``Vasculitis"] \\
Group 2: [``Cellulitis", ``Herpes HPV", ``Scabies Lyme", ``Tinea Ringworm"] \\
Group 3: [``Actinic", ``Melanoma"] \\
Group 4: [``Bullous", ``Seborrheic", ``Vascular Tumors"] \\
Group 5: [``Poison Ivy", ``Urticaria Hives"] \\
Group 6: [``Hair Loss", ``Nail Fungus"] \\
Group 7: [``Acne", ``Light Diseases", ``Lupus", ``Systemic", ``Warts Molluscum"] \\ \hline
13 classes \\\hline
Group 1: [``Atopic", ``Eczema", ``Psoriasis"] \\
Group 2: [``Cellulitis", ``Scabies Lyme"] \\
Group 3: [``Herpes HPV", ``Warts Molluscum"] \\
Group 4: [``Actinic", ``Melanoma"] \\
Group 5: [``Hair Loss"] \\
Group 6: [``Lupus", ``Vasculitis"] \\
Group 7: [``Light Diseases", ``Seborrheic"] \\
Group 8: [``Nail Fungus"] \\
Group 9: [``Exanthems", ``Poison Ivy"] \\
Group 10: [``Bullous", ``Systemic", ``Vascular Tumors"] \\
Group 11: [``Urticaria Hives"] \\
Group 12: [``Tinea Ringworm"] \\
Group 13: [``Acne"] \\
\hline
\end{tabular}
\end{table*}

\subsubsection{Creating Joint Feature Space}
\label{joint feature embedding}
In Section \ref{fisher CSS}, we measure CSS by mapping all datasets to a common feature space. Typically, this is achieved using the ImageNet pretrained model. However, this approach can introduce bias towards datasets closely aligned with ImageNet. Instead, we opted for a different approach: we used an ImageNet pretrained ResNet18 model and retrained it on all our datasets.

To achieve this, we combined the training sets of all datasets to create a larger dataset. It's important to note that each dataset has distinct training sizes and varying class numbers. Simply merging these imbalanced datasets for training would result in certain classes being underrepresented, given the differences in class sizes. To address this, we implemented a straightforward class-wise loss weighting technique to balance the training process. In this technique, we weighted the cross-entropy loss for each class by $\omega = \frac{\textit{No. of samples in the largest class}}{\textit{No. of samples in the given class}}$. Our training process employed a batch size of 512, utilized the SGD optimizer with a momentum of 0.9, weight decay of $10^{-4}$, and an initial learning rate of 0.05 for 50 epochs. We confirmed the effectiveness of this approach by evaluating the per-class F1-score on the test set, ensuring that none of the classes were overlooked.

\subsubsection{Class Grouping}
\label{class-grouping}

To group classes, as discussed in Section \ref{vsclass}, we merged similar classes to reduce the total number. COVID-DU-Ex required no changes due to its minimal class count, while other datasets were reorganized accordingly. NCT-CRC-HE-100K, originally with nine classes, was grouped into three, six, and seven using ChatGPT based on pathological conditions. MURA was categorized into three and six classes for anatomical proximity. DermNet's 23 classes were grouped into three, six, seven, and thirteen using ChatGPT. Maternal-fetal US was grouped into three classes: all fetal plane, maternal cervix, and others. Tables \ref{NCT-CRC-HE-100K-class-grouping}, \ref{Mura-class-grouping}, \ref{Dermnet-class-grouping}, and \ref{Fetal-class-grouping} illustrate the class groupings for NCT-CRC-HE-100K, MURA, DermNet, and Maternal-fetal US, respectively.

\subsubsection{Label Noise Injection}

In Section \ref{label noise injection}, we introduced both symmetrical and class-dependent label noise. Symmetrical label noise injection is straightforward (as discussed in Section \ref{label noise injection}), while for class-dependent label noise, we grouped classes within each dataset based on their relatedness to determine dependencies. Classes within a group were considered dependent, while any class outside its group was considered independent. For instance, in the COVID-DU-Ex dataset, both the ``Covid" and ``Non-covid" classes represent chest infections and are more likely to be mislabeled with each other than with ``Normal" condition. Consequently, these two classes were grouped together, allowing for the label of one to be altered to the other with a certain probability, while the "Normal" class remained unaffected. For other datasets, the groups are defined in the ``3 classes" column of Tables \ref{NCT-CRC-HE-100K-class-grouping}, \ref{Mura-class-grouping}, \ref{Fetal-class-grouping}, and \ref{Dermnet-class-grouping}.

A transition matrix depicts class dependencies through probability scores, with each ``row" representing the true class, and each ``column" indicating potential mislabeled classes. The probability score corresponds to the likelihood of a true class being changed to another class. When injecting class-dependent label noise, we only alter the label of a given class to those classes within the same group; the probability of flipping the label to any class outside the group is set to 0. Tables \ref{covid19_transition_matrix}, \ref{histopathology_transition_matrix}, \ref{mura_transition_matrix}, \ref{fetal_transition_matrix}, and \ref{dermnet_transition_matrix} show the transition matrices for each dataset. The probabilities depend on label noise rate $\epsilon$ and the count of candidate classes within a group, referred to as "spread" ($s$). Larger spreads indicate less severe label noise, reflecting greater class dependencies.
\begin{table}[!ht]
\scriptsize
\centering
\setlength{\tabcolsep}{4pt}
\caption{Transition Matrix for COVID-DU-Ex dataset, where $\epsilon$ denotes the label noise rate. The true classes are shown in \textbf{rows}, while the incorrect class are shown in \textbf{column}. Each element in the matrix depicts the probability of row class getting corrupted by corresponding column class.}
\label{covid19_transition_matrix}
\begin{tabular}{|l|l|l|l|}
\hline
          & Covid    & Non-Covid & Normal   \\ \hline
Covid     & 1-$\epsilon$ & $\epsilon$    & 0        \\ \hline
Non-Covid & $\epsilon$   & 1-$\epsilon$  & 0        \\ \hline
Normal    & 0        & 0         & 1-$\epsilon$ \\ \hline
\end{tabular}
\end{table}
\begin{table}[!ht]
\scriptsize
\centering
\setlength{\tabcolsep}{4pt}
\caption{Transition Matrix for Fetal dataset, where $\epsilon$ denotes the label noise rate. The true classes are shown in \textbf{rows}, while the incorrect class are shown in \textbf{column}. Each element in the matrix depicts the probability of row class getting corrupted by corresponding column class. ``$s$" denotes the spread, i.e the number of candidate classes with which the true class can be mislabeled.}
\label{fetal_transition_matrix}
\begin{tabular}{|l|l|l|l|l|l|l|}
\hline
                & Abdomen    & Brain      & Femur      & Thorax     & Cervix & Other \\ \hline
Abdomen   & 1-$\epsilon$         & $\frac{\epsilon}{s}$ & $\frac{\epsilon}{s}$ & $\frac{\epsilon}{s}$ & 0               & 0     \\ \hline
Brain     & $\frac{\epsilon}{s}$ & 1-$\epsilon$         & $\frac{\epsilon}{s}$ & $\frac{\epsilon}{s}$ & 0               & 0     \\ \hline
Femur     & $\frac{\epsilon}{s}$ & $\frac{\epsilon}{s}$ & 1-$\epsilon$         & $\frac{\epsilon}{s}$ & 0               & 0     \\ \hline
Thorax    & $\frac{\epsilon}{s}$ & $\frac{\epsilon}{s}$ & $\frac{\epsilon}{s}$ & 1-$\epsilon$         & 0               & 0     \\ \hline
Cervix & 0                & 0                & 0                & 0                & 0               & 0     \\ \hline
Other           & 0                & 0                & 0                & 0                & 0               & 0     \\ \hline
\end{tabular}
\end{table}
\begin{table}[!ht]
\scriptsize
\centering
\setlength{\tabcolsep}{4pt}
\caption{Transition Matrix for MURA dataset, where $\epsilon$ denotes the label noise rate. The true classes are shown in \textbf{rows}, while the incorrect class are shown in \textbf{column}. Each element in the matrix depicts the probability of row class getting corrupted by corresponding column class. ``$s_1$",``$s_2$", and ``$s_3$" denotes various spreads, i.e the number of candidate classes with which the true class can be mislabeled. }
\label{mura_transition_matrix}
\begin{tabular}{|l|l|l|l|l|l|l|l|}
\hline
         & Shoulder           & Humerus            & Forearm            & Finger             & Wrist              & Hand               & Elbow              \\ \hline
Shoulder & 1-$\epsilon$           & $\frac{\epsilon}{s_1}$ & 0                  & 0                  & 0                  & 0                  & 0                  \\ \hline
Humerus  & $\frac{\epsilon}{s_1}$ & 1-$\epsilon$           & 0                  & 0                  &                    & 0                  & 0                  \\ \hline
Forearm  & 0                  & 0                  & 1-$\epsilon$           & 0                  & $\frac{\epsilon}{s_3}$ & 0                  & $\frac{\epsilon}{s_3}$ \\ \hline
Finger   & 0                  & 0                  &       0             & 1-$\epsilon$           & 0                  & $\frac{\epsilon}{s_2}$ & 0                  \\ \hline
Wrist    & 0                  & 0                  & $\frac{\epsilon}{s_3}$ & 0                  & 1-$\epsilon$           & 0                  & $\frac{\epsilon}{s_3}$ \\ \hline
Hand     & 0                  & 0                  & 0                  & $\frac{\epsilon}{s_2}$ & 0                  & 1-$\epsilon$           & 0                  \\ \hline
Elbow    & 0                  & 0                  & $\frac{\epsilon}{s_3}$ & 0                  & $\frac{\epsilon}{s_3}$ & 0                  & 1-$\epsilon$           \\ \hline
\end{tabular}
\end{table}
\begin{table*}[!ht]
\scriptsize
\centering
\setlength{\tabcolsep}{4pt}
\caption{Transition Matrix for NCT-CRC-HE-100K dataset, where $\epsilon$ denotes the label noise rate. The true classes are shown in \textbf{rows}, while the incorrect class are shown in \textbf{column}. Each element in the matrix depicts the probability of row class getting corrupted by corresponding column class. ``$s_1$",``$s_2$", and ``$s_3$" denotes various spread, i.e the number of candidate classes with which the true class can be mislabeled.}
\label{histopathology_transition_matrix}
\begin{tabular}{|l|l|l|l|l|l|l|l|l|l|}
\hline
               & Adipose            & Smooth muscle      & Colon mucosa       & Background         & Debris             & Lymphocytes        & Mucus              & Cancer stroma      & Adenocarcinoma     \\ \hline
Adipose        & 1-$\epsilon$           & $\frac{\epsilon}{s_1}$ & $\frac{\epsilon}{s_1}$ & 0                  & 0                  & 0                  & 0                  & 0                  & 0                  \\ \hline
Smooth muscle  & $\frac{\epsilon}{s_1}$ & 1-$\epsilon$           & $\frac{\epsilon}{s_1}$ & 0                  & 0                  & 0                  & 0                  & 0                  & 0                  \\ \hline
Colon mucosa   & $\frac{\epsilon}{s_1}$ & $\frac{\epsilon}{s_1}$ & 1-$\epsilon$           & 0                  & 0                  & 0                  & 0                  & 0                  & 0                  \\ \hline
Background     & 0                  & 0                  & 0                  & 1-$\epsilon$           & $\frac{\epsilon}{s_2}$ & $\frac{\epsilon}{s_2}$ & $\frac{\epsilon}{s_2}$ & 0                  & 0                  \\ \hline
Debris         & 0                  & 0                  & 0                  & $\frac{\epsilon}{s_2}$ & 1-$\epsilon$           & $\frac{\epsilon}{s_2}$ & $\frac{\epsilon}{s_2}$ & 0                  & 0                  \\ \hline
Lymphocytes    & 0                  & 0                  & 0                  & $\frac{\epsilon}{s_2}$ & $\frac{\epsilon}{s_2}$ & 1-$\epsilon$           & $\frac{\epsilon}{s_2}$ & 0                  & 0                  \\ \hline
Mucus          & 0                  & 0                  & 0                  & $\frac{\epsilon}{s_2}$ & $\frac{\epsilon}{s_2}$ & $\frac{\epsilon}{s_2}$ & 1-$\epsilon$           & 0                  & 0                  \\ \hline
Cancer stroma  & 0                  & 0                  & 0                  & 0                  & 0                  & 0                  & 0                  & 1-$\epsilon$           & $\frac{\epsilon}{s_3}$ \\ \hline
Adenocarcinoma & 0                  & 0                  & 0                  & 0                  & 0                  & 0                  & 0                  & $\frac{\epsilon}{s_3}$ & 1-$\epsilon$           \\ \hline
\end{tabular}
\end{table*}
\begin{table*}[!ht]
\scriptsize
\centering
\setlength{\tabcolsep}{4pt}
\caption{Transition Matrix for Dermnet dataset, where $\epsilon$ denotes the label noise rate. The true classes are shown in \textbf{rows}, while the incorrect class are shown in \textbf{column}. Each element in the matrix depicts the probability of row class getting corrupted by corresponding column class. ``$s_1$",``$s_2$", and ``$s_3$" denotes various spreads, i.e the number of candidate classes with which the true class can be mislabeled. The corresponding class name for each class index is given in `` 3 classes" of Table \ref{Dermnet-class-grouping}, where index 1 denotes ``Acne",..,and 23 denotes ``Vasculitis"}
\label{dermnet_transition_matrix}
\begin{tabular}{|l|l|l|l|l|l|l|l|l|l|l|l|l|l|l|l|l|l|l|l|l|l|l|l|}
\hline
   & 1                  & 2                  & 3                  & 4                  & 5                  & 6                  & 7                  & 8                  & 9                  & 0                  & 11                 & 12                 & 13                 & 14                 & 15                 & 16                 & 17                 & 18                 & 19                 & 20                 & 21                 & 22                 & 23                 \\ \hline
1  & 1-$\epsilon$           & $\frac{\epsilon}{s_1}$ & $\frac{\epsilon}{s_1}$ & $\frac{\epsilon}{s_1}$ & $\frac{\epsilon}{s_1}$ & $\frac{\epsilon}{s_1}$ & $\frac{\epsilon}{s_1}$ & 0                  & 0                  & 0                  & 0                  & 0                  & 0                  & 0                  & 0                  & 0                  & 0                  & 0                  & 0                  & 0                  & 0                  & 0                  & 0                  \\ \hline
2  & $\frac{\epsilon}{s_1}$ & 1-$\epsilon$           & $\frac{\epsilon}{s_1}$ & $\frac{\epsilon}{s_1}$ & $\frac{\epsilon}{s_1}$ & $\frac{\epsilon}{s_1}$ & $\frac{\epsilon}{s_1}$ & 0                  & 0                  & 0                  & 0                  & 0                  & 0                  & 0                  & 0                  & 0                  & 0                  & 0                  & 0                  & 0                  & 0                  & 0                  & 0                  \\ \hline
3  & $\frac{\epsilon}{s_1}$ & $\frac{\epsilon}{s_1}$ & 1-$\epsilon$           & $\frac{\epsilon}{s_1}$ & $\frac{\epsilon}{s_1}$ & $\frac{\epsilon}{s_1}$ & $\frac{\epsilon}{s_1}$ & 0                  & 0                  & 0                  & 0                  & 0                  & 0                  & 0                  & 0                  & 0                  & 0                  & 0                  & 0                  & 0                  & 0                  & 0                  & 0                  \\ \hline
4  & $\frac{\epsilon}{s_1}$ & $\frac{\epsilon}{s_1}$ & $\frac{\epsilon}{s_1}$ & 1-$\epsilon$           & $\frac{\epsilon}{s_1}$ & $\frac{\epsilon}{s_1}$ & $\frac{\epsilon}{s_1}$ & 0                  & 0                  & 0                  & 0                  & 0                  & 0                  & 0                  & 0                  & 0                  & 0                  & 0                  & 0                  & 0                  & 0                  & 0                  & 0                  \\ \hline
5  & $\frac{\epsilon}{s_1}$ & $\frac{\epsilon}{s_1}$ & $\frac{\epsilon}{s_1}$ & $\frac{\epsilon}{s_1}$ & 1-$\epsilon$           & $\frac{\epsilon}{s_1}$ & $\frac{\epsilon}{s_1}$ & 0                  & 0                  & 0                  & 0                  & 0                  & 0                  & 0                  & 0                  & 0                  & 0                  & 0                  & 0                  & 0                  & 0                  & 0                  & 0                  \\ \hline
6  & $\frac{\epsilon}{s_1}$ & $\frac{\epsilon}{s_1}$ & $\frac{\epsilon}{s_1}$ & $\frac{\epsilon}{s_1}$ & $\frac{\epsilon}{s_1}$ & 1-$\epsilon$           & $\frac{\epsilon}{s_1}$ & 0                  & 0                  & 0                  & 0                  & 0                  & 0                  & 0                  & 0                  & 0                  & 0                  & 0                  & 0                  & 0                  & 0                  & 0                  & 0                  \\ \hline
7  & $\frac{\epsilon}{s_1}$ & $\frac{\epsilon}{s_1}$ & $\frac{\epsilon}{s_1}$ & $\frac{\epsilon}{s_1}$ & $\frac{\epsilon}{s_1}$ & $\frac{\epsilon}{s_1}$ & 1-$\epsilon$           & 0                  & 0                  & 0                  & 0                  & 0                  & 0                  & 0                  & 0                  & 0                  & 0                  & 0                  & 0                  & 0                  & 0                  & 0                  & 0                  \\ \hline
8  & 0                  & 0                  & 0                  & 0                  & 0                  & 0                  & 0                  & 1-$\epsilon$           & $\frac{\epsilon}{s_2}$ & $\frac{\epsilon}{s_2}$ & $\frac{\epsilon}{s_2}$ & 0                  & 0                  & 0                  & 0                  & 0                  & 0                  & 0                  & 0                  & 0                  & 0                  & 0                  & 0                  \\ \hline
9  & 0                  & 0                  & 0                  & 0                  & 0                  & 0                  & 0                  & $\frac{\epsilon}{s_2}$ & 1-$\epsilon$           & $\frac{\epsilon}{s_2}$ & $\frac{\epsilon}{s_2}$ & 0                  & 0                  & 0                  & 0                  & 0                  & 0                  & 0                  & 0                  & 0                  & 0                  & 0                  & 0                  \\ \hline
10 & 0                  & 0                  & 0                  & 0                  & 0                  & 0                  & 0                  & $\frac{\epsilon}{s_2}$ & $\frac{\epsilon}{s_2}$ & 1-$\epsilon$           & $\frac{\epsilon}{s_2}$ & 0                  & 0                  & 0                  & 0                  & 0                  & 0                  & 0                  & 0                  & 0                  & 0                  & 0                  & 0                  \\ \hline
11 & 0                  & 0                  & 0                  & 0                  & 0                  & 0                  & 0                  & $\frac{\epsilon}{s_2}$ & $\frac{\epsilon}{s_2}$ & $\frac{\epsilon}{s_2}$ & 1-$\epsilon$           & 0                  & 0                  & 0                  & 0                  & 0                  & 0                  & 0                  & 0                  & 0                  & 0                  & 0                  & 0                  \\ \hline
12 & 0                  & 0                  & 0                  & 0                  & 0                  & 0                  & 0                  & 0                  & 0                  & 0                  & 0                  & 1-$\epsilon$           & $\frac{\epsilon}{s_3}$ & $\frac{\epsilon}{s_3}$ & $\frac{\epsilon}{s_3}$ & $\frac{\epsilon}{s_3}$ & $\frac{\epsilon}{s_3}$ & $\frac{\epsilon}{s_3}$ & $\frac{\epsilon}{s_3}$ & $\frac{\epsilon}{s_3}$ & $\frac{\epsilon}{s_3}$ & $\frac{\epsilon}{s_3}$ & $\frac{\epsilon}{s_3}$ \\ \hline
13 & 0                  & 0                  & 0                  & 0                  & 0                  & 0                  & 0                  & 0                  & 0                  & 0                  & 0                  & $\frac{\epsilon}{s_3}$ & 1-$\epsilon$           & $\frac{\epsilon}{s_3}$ & $\frac{\epsilon}{s_3}$ & $\frac{\epsilon}{s_3}$ & $\frac{\epsilon}{s_3}$ & $\frac{\epsilon}{s_3}$ & $\frac{\epsilon}{s_3}$ & $\frac{\epsilon}{s_3}$ & $\frac{\epsilon}{s_3}$ & $\frac{\epsilon}{s_3}$ & $\frac{\epsilon}{s_3}$ \\ \hline
14 & 0                  & 0                  & 0                  & 0                  & 0                  & 0                  & 0                  & 0                  & 0                  & 0                  & 0                  & $\frac{\epsilon}{s_3}$ & $\frac{\epsilon}{s_3}$ & 1-$\epsilon$           & $\frac{\epsilon}{s_3}$ & $\frac{\epsilon}{s_3}$ & $\frac{\epsilon}{s_3}$ & $\frac{\epsilon}{s_3}$ & $\frac{\epsilon}{s_3}$ & $\frac{\epsilon}{s_3}$ & $\frac{\epsilon}{s_3}$ & $\frac{\epsilon}{s_3}$ & $\frac{\epsilon}{s_3}$ \\ \hline
15 & 0                  & 0                  & 0                  & 0                  & 0                  & 0                  & 0                  & 0                  & 0                  & 0                  & 0                  & $\frac{\epsilon}{s_3}$ & $\frac{\epsilon}{s_3}$ & $\frac{\epsilon}{s_3}$ & 1-$\epsilon$           & $\frac{\epsilon}{s_3}$ & $\frac{\epsilon}{s_3}$ & $\frac{\epsilon}{s_3}$ & $\frac{\epsilon}{s_3}$ & $\frac{\epsilon}{s_3}$ & $\frac{\epsilon}{s_3}$ & $\frac{\epsilon}{s_3}$ & $\frac{\epsilon}{s_3}$ \\ \hline
16 & 0                  & 0                  & 0                  & 0                  & 0                  & 0                  & 0                  & 0                  & 0                  & 0                  & 0                  & $\frac{\epsilon}{s_3}$ & $\frac{\epsilon}{s_3}$ & $\frac{\epsilon}{s_3}$ & $\frac{\epsilon}{s_3}$ & 1-$\epsilon$           & $\frac{\epsilon}{s_3}$ & $\frac{\epsilon}{s_3}$ & $\frac{\epsilon}{s_3}$ & $\frac{\epsilon}{s_3}$ & $\frac{\epsilon}{s_3}$ & $\frac{\epsilon}{s_3}$ & $\frac{\epsilon}{s_3}$ \\ \hline
17 & 0                  & 0                  & 0                  & 0                  & 0                  & 0                  & 0                  & 0                  & 0                  & 0                  & 0                  & $\frac{\epsilon}{s_3}$ & $\frac{\epsilon}{s_3}$ & $\frac{\epsilon}{s_3}$ & $\frac{\epsilon}{s_3}$ & $\frac{\epsilon}{s_3}$ & 1-$\epsilon$           & $\frac{\epsilon}{s_3}$ & $\frac{\epsilon}{s_3}$ & $\frac{\epsilon}{s_3}$ & $\frac{\epsilon}{s_3}$ & $\frac{\epsilon}{s_3}$ & $\frac{\epsilon}{s_3}$ \\ \hline
18 & 0                  & 0                  & 0                  & 0                  & 0                  & 0                  & 0                  & 0                  & 0                  & 0                  & 0                  & $\frac{\epsilon}{s_3}$ & $\frac{\epsilon}{s_3}$ & $\frac{\epsilon}{s_3}$ & $\frac{\epsilon}{s_3}$ & $\frac{\epsilon}{s_3}$ & $\frac{\epsilon}{s_3}$ & 1-$\epsilon$           & $\frac{\epsilon}{s_3}$ & $\frac{\epsilon}{s_3}$ & $\frac{\epsilon}{s_3}$ & $\frac{\epsilon}{s_3}$ & $\frac{\epsilon}{s_3}$ \\ \hline
19 & 0                  & 0                  & 0                  & 0                  & 0                  & 0                  & 0                  & 0                  & 0                  & 0                  & 0                  & $\frac{\epsilon}{s_3}$ & $\frac{\epsilon}{s_3}$ & $\frac{\epsilon}{s_3}$ & $\frac{\epsilon}{s_3}$ & $\frac{\epsilon}{s_3}$ & $\frac{\epsilon}{s_3}$ & $\frac{\epsilon}{s_3}$ & 1-$\epsilon$           & $\frac{\epsilon}{s_3}$ & $\frac{\epsilon}{s_3}$ & $\frac{\epsilon}{s_3}$ & $\frac{\epsilon}{s_3}$ \\ \hline
20 & 0                  & 0                  & 0                  & 0                  & 0                  & 0                  & 0                  & 0                  & 0                  & 0                  & 0                  & $\frac{\epsilon}{s_3}$ & $\frac{\epsilon}{s_3}$ & $\frac{\epsilon}{s_3}$ & $\frac{\epsilon}{s_3}$ & $\frac{\epsilon}{s_3}$ & $\frac{\epsilon}{s_3}$ & $\frac{\epsilon}{s_3}$ & $\frac{\epsilon}{s_3}$ & 1-$\epsilon$           & $\frac{\epsilon}{s_3}$ & $\frac{\epsilon}{s_3}$ & $\frac{\epsilon}{s_3}$ \\ \hline
21 & 0                  & 0                  & 0                  & 0                  & 0                  & 0                  & 0                  & 0                  & 0                  & 0                  & 0                  & $\frac{\epsilon}{s_3}$ & $\frac{\epsilon}{s_3}$ & $\frac{\epsilon}{s_3}$ & $\frac{\epsilon}{s_3}$ & $\frac{\epsilon}{s_3}$ & $\frac{\epsilon}{s_3}$ & $\frac{\epsilon}{s_3}$ & $\frac{\epsilon}{s_3}$ & $\frac{\epsilon}{s_3}$ & 1-$\epsilon$           & $\frac{\epsilon}{s_3}$ & $\frac{\epsilon}{s_3}$ \\ \hline
22 & 0                  & 0                  & 0                  & 0                  & 0                  & 0                  & 0                  & 0                  & 0                  & 0                  & 0                  & $\frac{\epsilon}{s_3}$ & $\frac{\epsilon}{s_3}$ & $\frac{\epsilon}{s_3}$ & $\frac{\epsilon}{s_3}$ & $\frac{\epsilon}{s_3}$ & $\frac{\epsilon}{s_3}$ & $\frac{\epsilon}{s_3}$ & $\frac{\epsilon}{s_3}$ & $\frac{\epsilon}{s_3}$ & $\frac{\epsilon}{s_3}$ & 1-$\epsilon$           & $\frac{\epsilon}{s_3}$ \\ \hline
23 & 0                  & 0                  & 0                  & 0                  & 0                  & 0                  & 0                  & 0                  & 0                  & 0                  & 0                  & $\frac{\epsilon}{s_3}$ & $\frac{\epsilon}{s_3}$ & $\frac{\epsilon}{s_3}$ & $\frac{\epsilon}{s_3}$ & $\frac{\epsilon}{s_3}$ & $\frac{\epsilon}{s_3}$ & $\frac{\epsilon}{s_3}$ & $\frac{\epsilon}{s_3}$ & $\frac{\epsilon}{s_3}$ & $\frac{\epsilon}{s_3}$ & $\frac{\epsilon}{s_3}$ & 1-$\epsilon$           \\ \hline
\end{tabular}
\end{table*}

\subsection{Implementation Details for Proposed Pipeline}

In this section, we delve into the implementation details of self-supervised learning methods across all datasets, as well as the supervised LNL methods employed for medical image classification with noisy labels. Throughout our experiments, we utilized ResNet18 as the backbone.

\subsubsection{Self-supervised Pretraining}
\begin{itemize}[leftmargin=*]
    \item \textbf {Pretext tasks: }
The rotation prediction task involved predicting the rotation angle of the input image. To train for the rotation prediction task, we preprocessed the input image by applying strong data augmentations randomly, which included horizontal flips, slight rotations (within a range of $10^{\circ}$), sharpness adjustment, equalization, and auto-contrast. We trained to predict four rotation angles: $0^{\circ}$, $90^{\circ}$, $180^{\circ}$, and $270^{\circ}$.

Similarly, the Jigsaw Puzzle solving task also incorporated strong augmentations, mirroring those used in the rotation prediction task. After these augmentations, the input image was partitioned into a $3\times3$ grid of patches. We resized all the patches to $64 \times 64$ pixels and then normalized each patch individually using its patch mean and standard deviation. For each image, nine patches were randomly shuffled to create a unique permutation. Following the approach outlined in \footnote{https://github.com/bbrattoli/JigsawPuzzlePytorch}, we generated a total of 1000 distinct permutations. These patches were subsequently passed through ResNet18 for feature extraction. The extracted features were then concatenated into a single vector, which was fed into a fully connected layer comprising 1000 neurons to predict one of the 1000 possible permutations.

The JigMag puzzle solving task followed precisely the same preprocessing approach as the Jigsaw puzzle solving task. However, the grid of $3\times3$ patches was formed by arranging nine patches obtained after randomly magnifying different locations of the image using nine magnification factors ranging from 1 to 2.25, incremented by 0.25. Prior to feeding into the ResNet18 feature extractor, these patches were self-normalized using local patch mean and standard deviation. The resulting features were concatenated to form a long vector, which was then passed through a fully connected layer with 1000 neurons for predicting from among 1000 chosen permutations.

\item \textbf{Contrastive Learning: }
We trained SimCLR using this GitHub implementation\footnote{https://github.com/sthalles/SimCLR}, that uses a 2-layer MLP head (512-d, ReLU, 128-d) at the end layer, after ResNet18 base encoder. We followed the random augmentations used in the original paper, i.e cropping and resize to the original size, horizontal flips, color distortions, and Gaussian blur. For MoCo, we used the GitHub official implementation\footnote{https://github.com/facebookresearch/moco} that offers MoCo v2 \citep{chen2020improved}, which is an improved version that uses additional augmentations and a MLP projection head as in SimCLR. We used the default queue size (K = 65536) and momentum for encoder (m = 0.999). For BarlowTwin, we adapted the official GitHub implementation\footnote{https://github.com/facebookresearch/barlowtwins} that uses a 3-layer MLP projection head. We used the default 8192-d projection head.

\item \textbf{Generative Approaches: }
We adapted this GitHub repository\footnote{https://github.com/hsinyilin19/ResNetVAE} for the VAE. We  replaced the original encoder with a ResNet18 backbone, supplemented by a 3-layer MLP that outputs 256-dimensional mean (\(\mu\)) and variance (\(\varepsilon\)) vectors. The decoder consists of a 2-layer MLP, followed by three up-sampling transpose convolutional layers and a final bilinear interpolator. We set the weight for the KLD loss, \(\beta\), to 0.1. For the BigBiGAN implementation, we followed this GitHub repository\footnote{https://github.com/RKorzeniowski/BigBiGAN-PyTorch}. In this case, the encoder utilizes a ResNet18 backbone followed by an MLP. We configured the latent vector to have 114 dimensions and adjusted the generator channels to produce a \(128 \times 128\) output. During the training process, we set the betas for the Adam optimizer to (0.5, 0.999) and performed two updates for the discriminator for every update of the encoder and the generator.

General hyperparameters are provided in Table \ref{self-supervised hyperparameter general}. Training epochs are specific to each method and dataset and are reported in Table \ref{self-supervised hyperparameter dataset specific}. We trained all methods until convergence, upto the epochs where the training curve begins to saturate. For generative models, we selected the best model through qualitative inspection of the generated images at each epoch. In some cases, prolonging the training duration negatively impacted the quality of generation, as observed in both VAE and BigBiGAN. Therefore, visual inspection of the generated images was essential for choosing the best epoch model.

\end{itemize}
\begin{table}[!ht]
\scriptsize
\centering
\setlength{\tabcolsep}{4pt}
\caption{Method-specific general hyperparameters common across all the datasets.}
\label{self-supervised hyperparameter general}
\begin{tabular}{l|l|l|l|l|l|l}
\hline
 Method & Input size    & Batch & Wt decay & Lr & Optim & Sched        \\ \hline
 Rotation                                                        & $224 \times 224$     & $256$         & $10^{-4}$         & $0.01$                                                             & SGD       & Cos\\ \hline 
                                  Jigsaw                                                          & $64 \times 64$ & $128$            & $0$         & $0.001$                                                            & Adam      & Cos \\ \hline
                                  Jigmag                                                          & $64 \times 64$ & $128$        & $0$         & $0.0001$                                                            & Adam      & Cos \\ \hline 
                                  SimCLR                                                          & $224 \times 224$     & $512$         & $10^{-4}$         & $0.001$                                                            & Adam      & Cos \\ \hline 

                                  BarlowTwin                                                          & $224 \times 224$     & 512          & $10^{-6}$         & $0.2$                                                            & LARS      & Cos \\ \hline 

                                  Moco v2                                                          & $224 \times 224$     & 512           & $10^{-4}$         & $0.01$                                                            & SGD      & Cos\\ \hline 
                                  VAE                                                          & $224 \times 224$     & 512          & $0$         & $0.001$                                                            & Adam      & Cos\\ \hline 
                                  BigBiGAN                                                          & $128 \times 128$     & $32$         & $0$    & $2e^{-5}$                                                                  & Adam      & - \\ \hline
                                 
\end{tabular}
\end{table}



\begin{table}[!ht]
\scriptsize
\centering
\setlength{\tabcolsep}{4pt}
\caption{Training epochs for each datasets across all self-supervised methods}
\label{self-supervised hyperparameter dataset specific}
\begin{tabular}{c|c|c|c|c|c}
\hline
Method & \begin{tabular}[c]{@{}c@{}}NCT-CRC-\\ H-100K\end{tabular} & \begin{tabular}[c]{@{}c@{}}COVID-\\ QU-Ex\end{tabular}& MURA & DERMNET & FETAL \\ \hline
Rotation & $150$ & $150$& $150$ & $150$& $150$ \\ \hline
Jigsaw & $300$& $300$& $300$& $300$& $300$ \\ \hline
Jigmag & $300$& $300$& $300$& $300$& $300$ \\ \hline
SimCLR & $400$& $600$& $600$& $600$& $800$ \\ \hline
BarlowTwin& $300$& $600$& $600$& $600$& $800$ \\ \hline
Moco v2 & $500$& $500$& $600$& $600$& $700$ \\ \hline
VAE & $200$& $100$ & $49$& $400$& $400$ \\ \hline
BigBiGAN & $30$& $250$& $100$& $200$& $350$ \\ \hline
\end{tabular}
\end{table}

\subsubsection{Supervised Learning with LNL}
\label{learning with noisy labels hyperparameters}
We first trained ResNet18 from scratch on respective noisy medical classification datasets using standard cross-entropy (CE), Co-teaching (CT) and Dividemix (DM) to establish a baseline. For training, we used the SGD optimizer with an initial learning rate of 0.01, momentum of 0.9, and weight decay of $10^{-4}$. Standard cross-entropy and CT used a batch size of 256, while DM performed best at 128. All datasets were trained for 50 epochs, the point at which the training saturated, except for DermNet, which required 100 epochs. We conducted tests using both approaches with symmetrical label noise for CT to determine the best method. For both CT and DM, we mostly used method-specific hyperparameters from the original papers with some adjustments. For CT, we set $\tau = \epsilon$, where $\epsilon$ is the given label noise rate, and $c = 1$ for all datasets. For DM, we set $M = 2$ and $\alpha = 4$. $\lambda_{u} = 0$ was for $p = {0.5, 0.6, 0.7}$, but $\lambda_{u} = 25$. For all datasets, we used $T = 0.2$, except for Dermnet, for which we used $T = 0.5$. Remember that both methods used the same warm-up epochs of 10. All LNL experiments were run three times with different seeds to report the mean and standard deviation. We used a 40 GB A100 GPU to run both self-supervised pretraining and LNL experiments using PyTorch 1.2.1 on Python 3.8. 

Then, we initialized the ResNet18 backbone with self-supervised pretrained weights and fine-tuned it on respective datasets. It should be noted that here are two potential approaches for adapting the pretrained model: i) plastic backbone, where the weights of the entire pretrained backbone are trainable, and ii) frozen backbone, where most of the pretrained model's layers remain frozen. For the frozen backbone strategy, all 16 CNN layers of the ResNet18 backbone were frozen, except the last two convolutional layers. After reviewing the initial results, we chose to use a plastic backbone for all subsequent experiments.

\subsection{Evaluation}
We evaluated the classification performance using the F1-score, which captures both the Recall and Precision. Since the test set is not perfectly balanced, we computed the macro-average of the per class F1-score to measure the average F1-score. An important criterion to assess the robustness against noisy labels is to measure if the model overfits the noisy labels over the training. Therefore, we measure the average of the test F1-score in the last five epochs, indicated by LAST. In addition, we also measure the best F1-score achieved in the test, indicated by BEST, to assess the maximum performance.

\section{Results}
\label{results}
\subsection{Symmetrical Label Noise}
\begin{itemize}[leftmargin=*]
    \item \textbf{Plastic Backbone: }
We first compare all the methods by keeping self-supervised pretrained backbone plastic while training the supervised LNL of second phase. In this section, we only focus on symmetrical label noise. We show the test F1-score across five datasets in Fig. \ref{fig:symmetrical_coteaching_plastic_backbone_details} and Fig. \ref{fig:symmetrical_dividemix_plastic_backbone_detials}. The standard cross-entropy (CE) serves as the absolute baseline that does not employ any LNL approach. CE, and both the original Co-teaching (CT) and Dividemix (DM) are trained from scratch, while the term before ``+' indicates a pretrained model—either from generative, pretext task-based, or contrastive learning-based approaches.  We separately compare all self-supervised methods across CT and DM. In addition to self-supervised pretraining, we also compare it with the ImageNet pretrained model. Of all the compared approaches, contrastive learning-based pretraining (MoCo V2, BarlowTwin, SimCLR) performs the best in all datasets for both CT and DM. We observed that both generative approaches (VAE, BigBiGAN) performed poorly compared to the original CT and DM, likely because both are trained with an image construction objective, not suitable for downstream classification tasks. For pretext task-based approaches, only the Rotation task improved the performance of CT and DM, while Jigsaw and Jigmag performed poorly. In Table \ref{best_self_supervised_method}, we present the best pretraining method for each category of approaches from a total of eight self-supervised methods. Additionally, we also compare the test confusion matrix of CT trained from scratch against CT with best pretrained model in Fig. \ref{fig:confusion_matrix}. The confusion matrix of CT with a pretrained model across all datasets exhibits fewer off-diagonal values compared to CT trained from scratch. Especially with COVID-DU-Ex and DermNet, the performance is poor without pretraining, as depicted by the respective confusion matrices where some class labels are completely missed in predictions. Pretraining completely eliminates this issue with COVID-DU-Ex and significantly minimizes it in DermNet. This observation underscores the effectiveness of pretraining in enhancing robustness against noisy labels and, consequently, improving per-class predictions.

\begin{figure}[h!]
\centering
\includegraphics[width=1\linewidth]{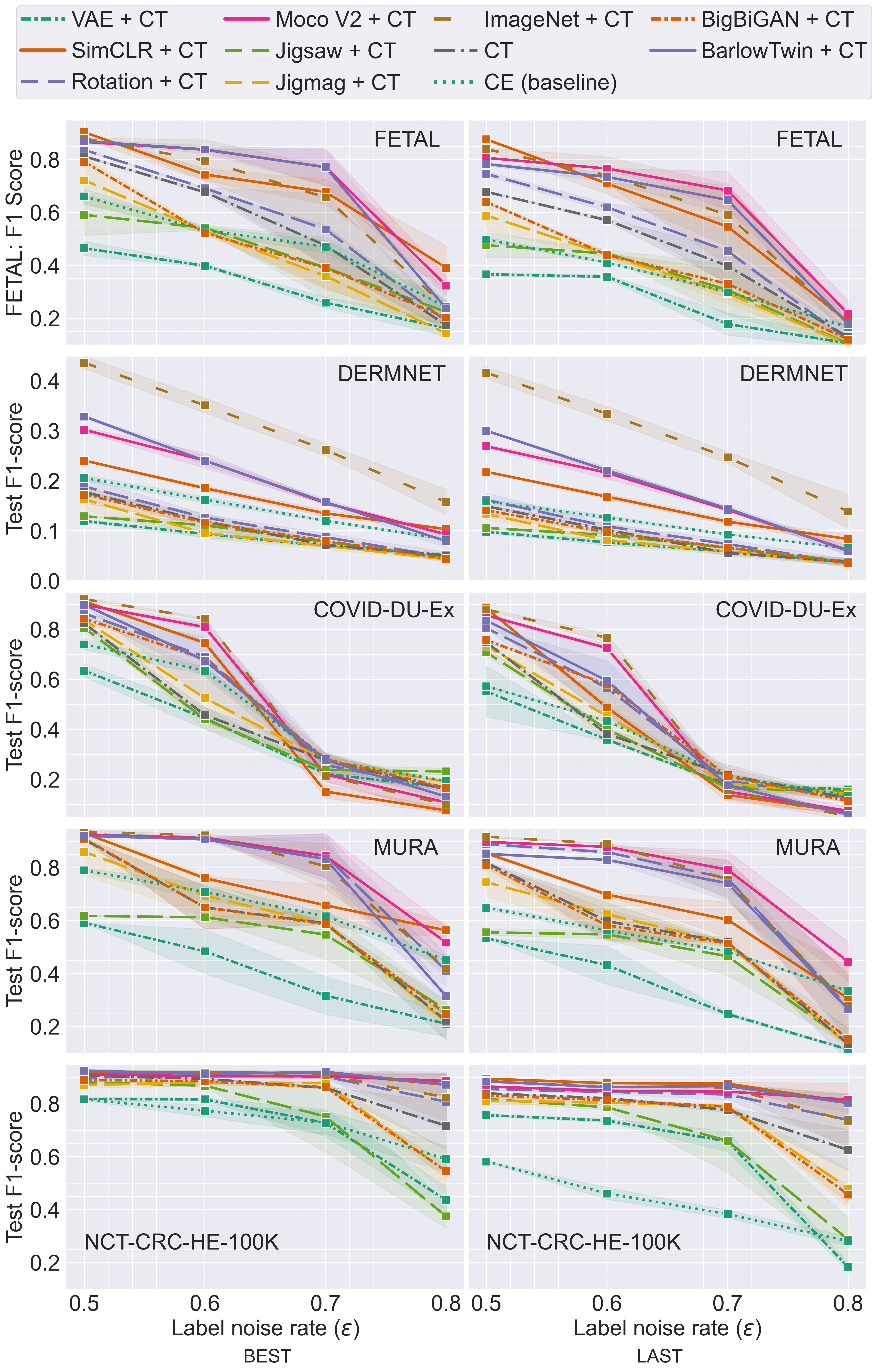}
   
\vspace{-0.5em}
    \caption{Comparing the test performance of Co-teaching (CT) with various self-supervised pretraining (\textbf{plastic backbone}) at different \textbf{symmetrical} label noise rates, across five datasets. CE stands for standard cross-entropy. LAST and BEST show the best performance and average of the last five epochs, respectively in the test set.}
     \label{fig:symmetrical_coteaching_plastic_backbone_details}
\end{figure}
  
\begin{figure}[h!]
    \includegraphics[width=1\linewidth]{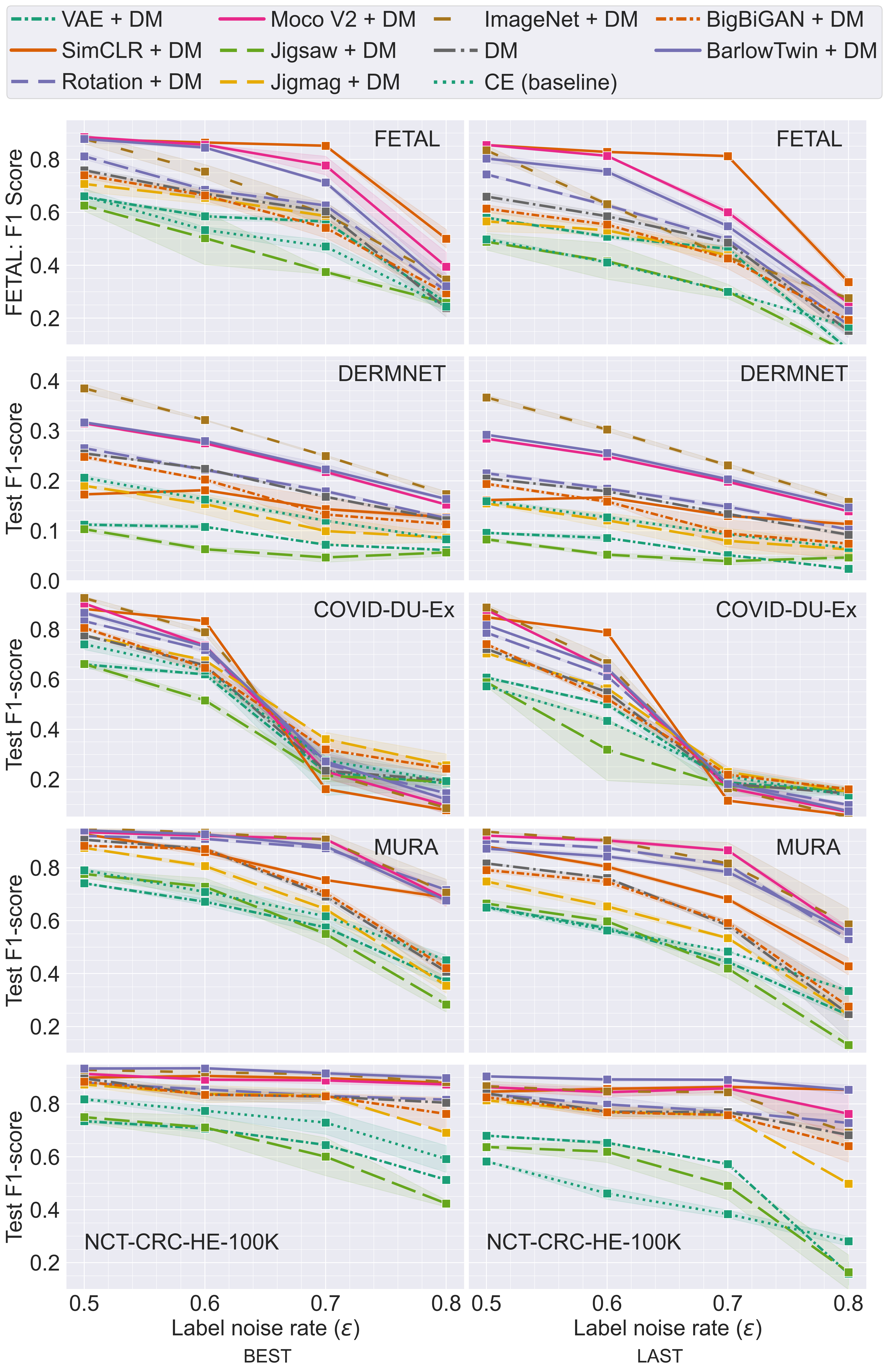}
   
    \caption{Comparing the test performance of Dividemix (DM) with various self-supervised pretraining (\textbf{plastic backbone}) at different \textbf{symmetrical} label noise rates, across five datasets. CE stands for standard cross-entropy. LAST and BEST show the best performance and average of the last five epochs, respectively in the test set.}
     \label{fig:symmetrical_dividemix_plastic_backbone_detials}
\end{figure}

\begin{figure}[h!]
    \includegraphics[width=1\linewidth]{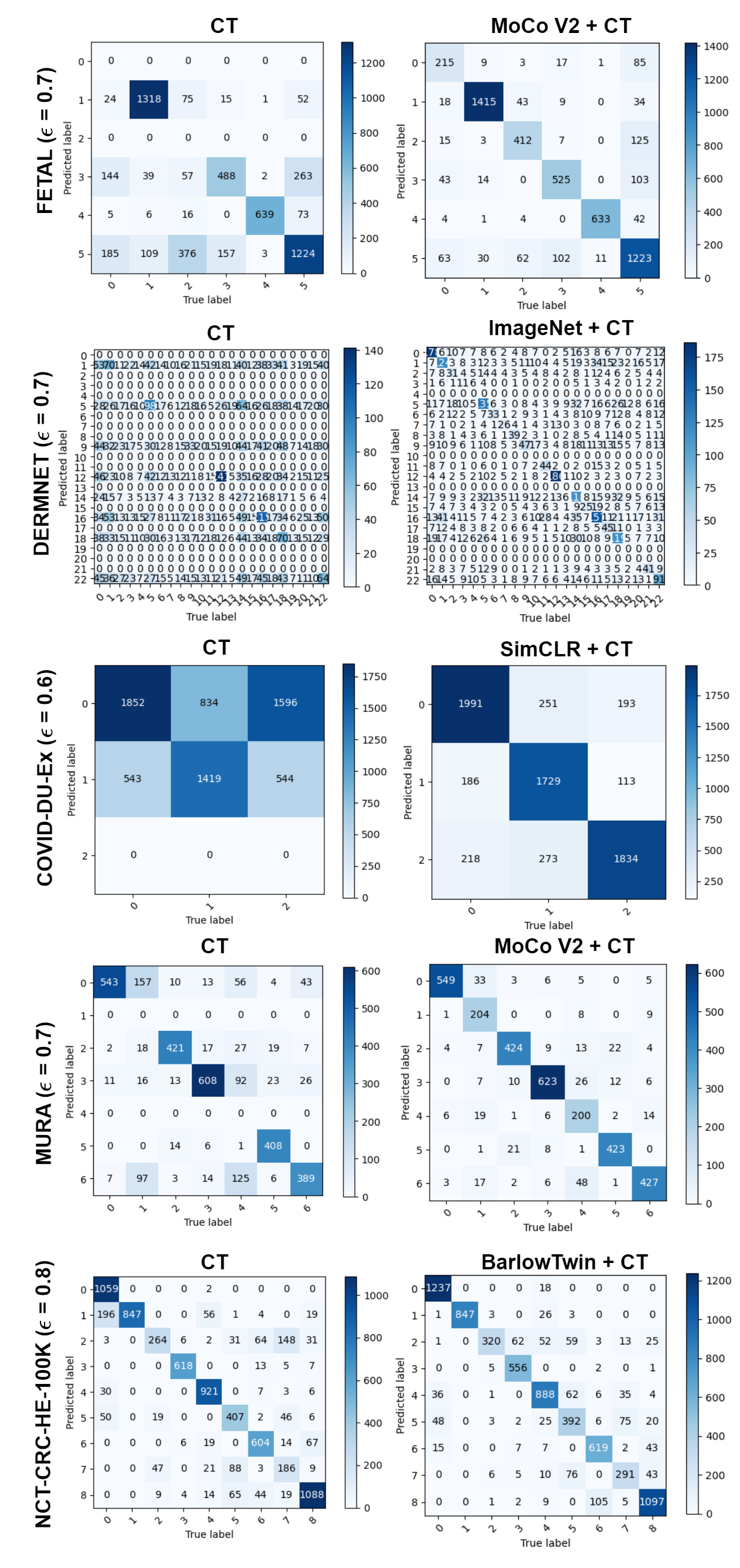}
    \caption{Comparing confusion matrix of Co-teaching (CT) with best pretrained model (\textbf{plastic backbone}) across various datasets at respective symmetrical noise rates ($\epsilon$).}
    \label{fig:confusion_matrix}
\end{figure}

\begin{table}[h!]
\scriptsize
\centering
\setlength{\tabcolsep}{4pt}
\caption{Best method from pretext task-based, contrastive learning-based, and generative approaches across all the datasets. Out of three, contrastive learning-based pretraining is most robust against noisy labels.}
\label{best_self_supervised_method}
\begin{tabular}{c|c|c|c|c|c}
\hline
Method & \begin{tabular}[c]{@{}c@{}}NCT-CRC-\\ H-100K\end{tabular} & \begin{tabular}[c]{@{}c@{}}COVID-\\ QU-Ex\end{tabular}& MURA & DERMNET & FETAL \\ \hline
Pretext& Rotation & Rotation & Rotation & Rotation & Rotation  \\ \hline
Contrastive& BarlowTwin & \begin{tabular}[c]{@{}c@{}}MoCo v2/\\ SimCLR\end{tabular}& MoCo v2 & BarlowTwin  & \begin{tabular}[c]{@{}c@{}}MoCo v2/\\ SimCLR\end{tabular}  \\ \hline
Generative&BigBiGAN& BigBiGAN&BigBiGAN&BigBiGAN&BigBiGAN \\\hline
\end{tabular}
\end{table}

Interestingly, we found that the ImageNet pretrained model, despite being trained on an out-of-domain dataset, performed relatively better, particularly in DermNet where it outperformed in-domain self-supervised pretraining. Two possible reasons for this are: i) the dataset size of DermNet is relatively smaller than NCT-CRC-HE-100K and MURA, limiting the full potential of self-supervised learning, and ii) the dataset contains RGB camera-captured images, whose features align closely with ImageNet images. Supporting the second reason, in-domain contrastive learning-based pretraining was superior to the ImageNet pretrained model despite having a smaller dataset size, as the Fetal dataset contains ultrasound images with a greater domain gap from ImageNet images.

\textit{In summary, we observed that contrastive learning-based is the best approach to pretrain the model for improving robustness against noisy labels in medical image classification, and self-supervised training with in-domain data is better than the ImageNet pretrained model if the medical images have a large domain gap with ImageNet or have a larger dataset size.}

\item \textbf{Frozen Backbone: }

We also investigated how freezing the backbone would perform by training CT with symmetrical label noise by freezing most layers of pretrained model (shown in Fig. \ref{fig:symmetrical_coteaching_frozen_backbone_details}). \textit{We observed that keeping backbone plastic is a better approach than freezing the layers for all the datasets}. It is important to keep the backbone plastic especially when image input preprocessing for pretraining differs significantly from that of downstream LNL classification tasks. Especially Jigsaw, Jigmag, VAE and BigBiGAN yielded the worst results when layers were frozen as these methods employ slightly different preprocessing steps.

\begin{figure}[h!]
\centering
\includegraphics[width=1\linewidth]{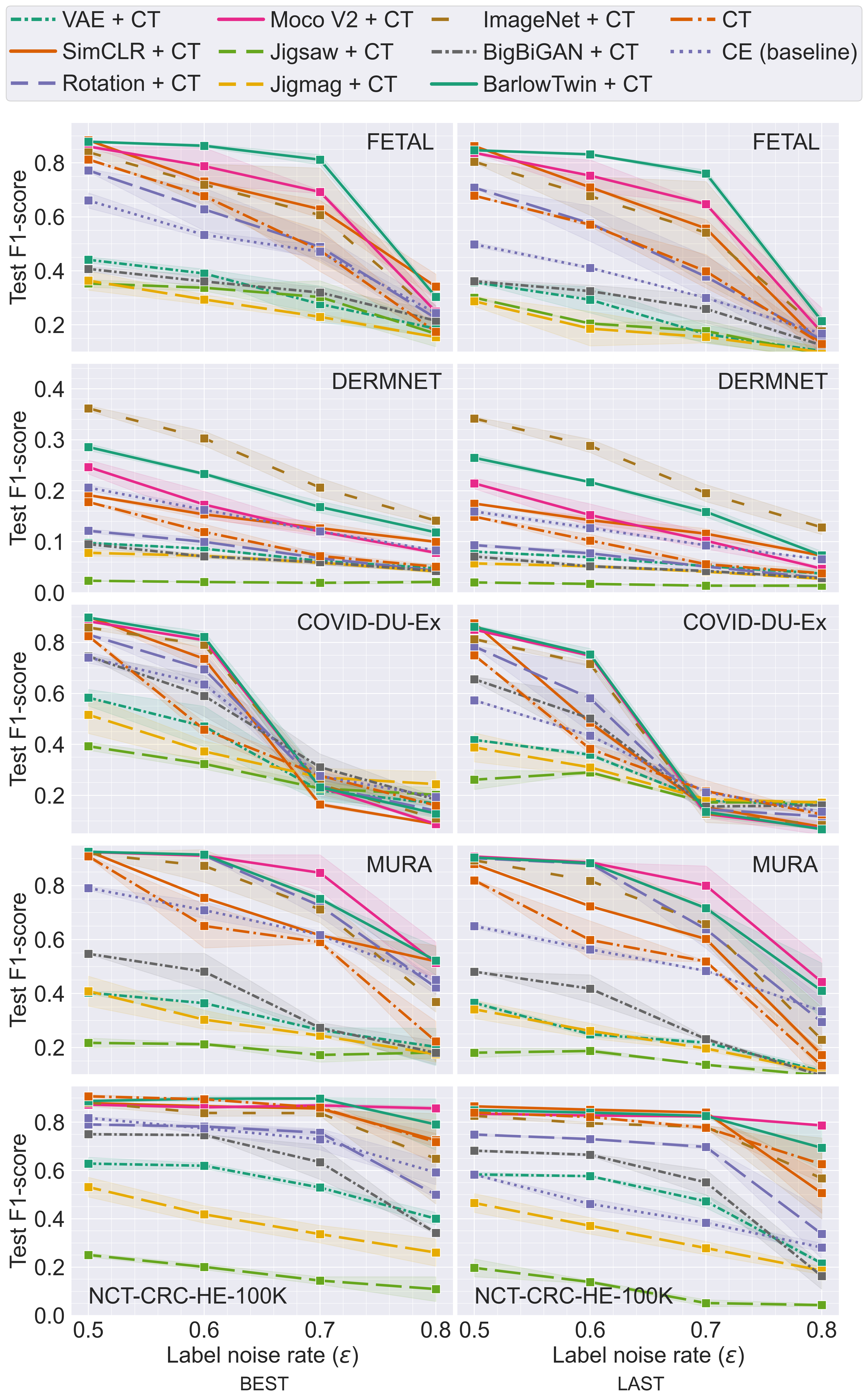}
   
\vspace{-0.5em}
    \caption{Comparing the test performance of Co-teaching (CT) with various self-supervised pretraining (\textbf{frozen backbone}) at different \textbf{symmetrical} label noise rates, across five datasets. CE stands for standard cross-entropy. LAST and BEST show the best performance and average of the last five epochs, respectively in the test set.}
     \label{fig:symmetrical_coteaching_frozen_backbone_details}
\end{figure}

\end{itemize}

\subsection{Class-dependent Label Noise}
\label{class-dependent label noise}
\begin{figure}[h!]
\centering
\includegraphics[width=1\linewidth]{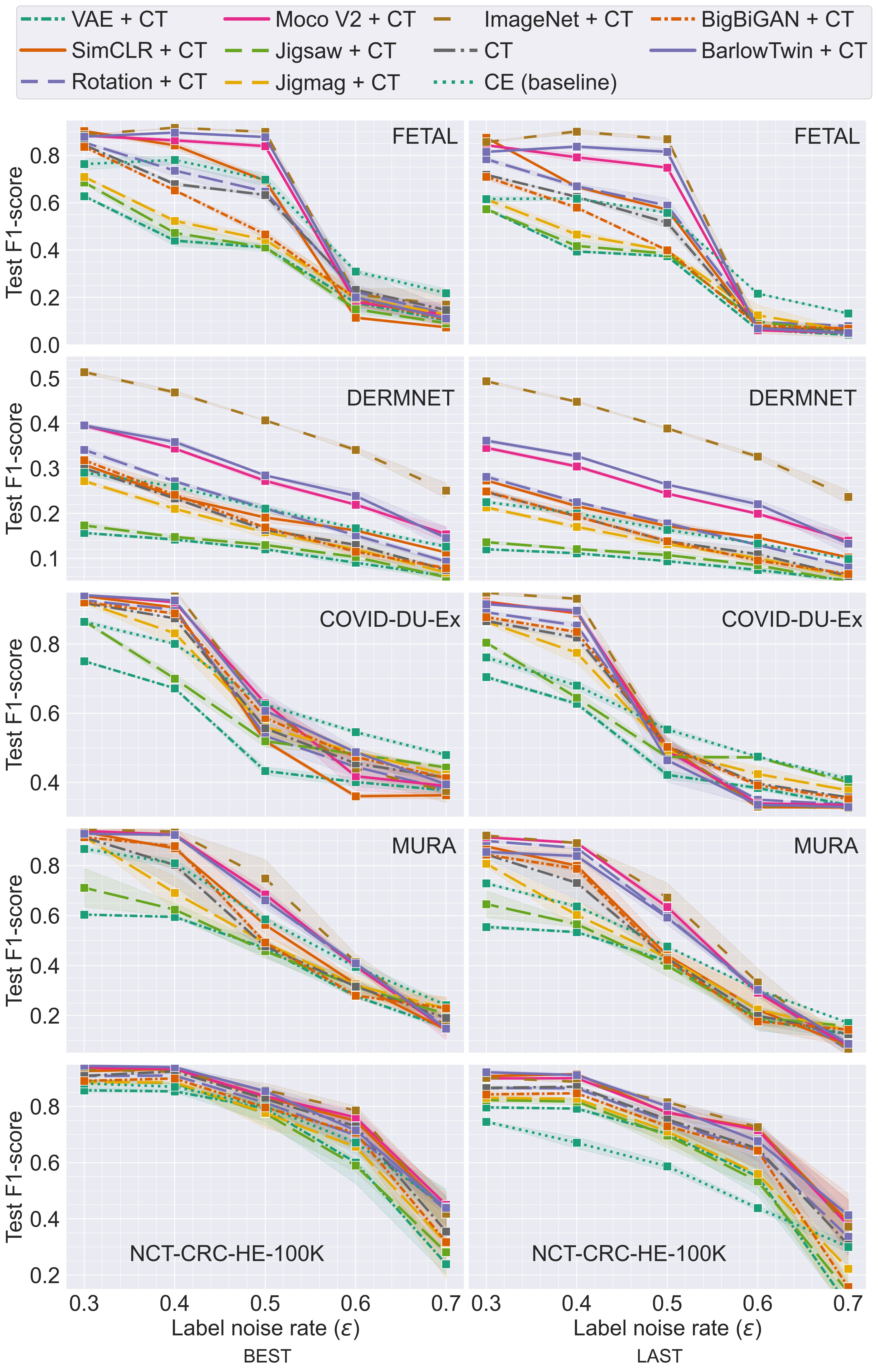}
    
\vspace{-0.5em}
    \caption{Comparing the test performance of Co-teaching (CT) with various self-supervised pretraining (\textbf{plastic backbone}) at different \textbf{class-dependent} label noise rates, across five datasets. CE stands for standard cross-entropy. LAST and BEST show the best performance and average of the last five epochs, respectively in the test set.}
    \label{fig:class_dependent_coteaching_plastic_backbone_detials}
\end{figure}

\begin{figure}[h!]
    \includegraphics[width=1\linewidth]{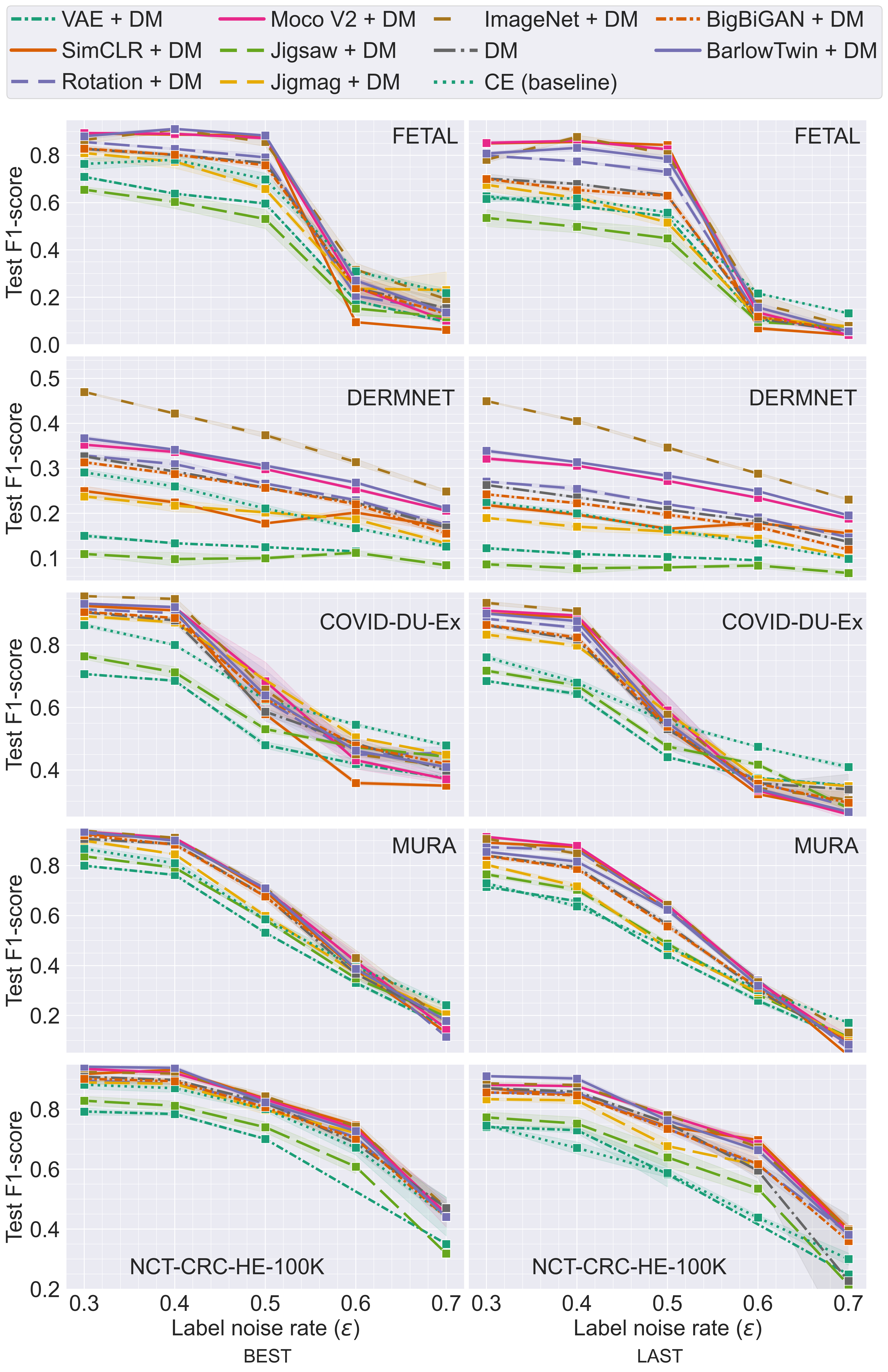}
    
    \caption{Comparing the test performance of Dividemix (DM) with various self-supervised pretraining (\textbf{plastic backbone}) at different \textbf{class-dependent} label noise rates, across five datasets. CE stands for standard cross-entropy. LAST and BEST show the best performance and average of the last five epochs, respectively in the test set.}
    \label{fig:class_dependent_dividemix_plastic_backbone_details}
\end{figure}

Finally, we also investigated all self-supervised pretraining with class-dependent label noise. As depicted in Section \ref{class-dependent-theory}, class-dependent label noise is theoretically more detrimental than symmetrical label noise, as the flipping threshold is relatively very low. Under these circumstances, even good features cannot guarantee improved robustness. Our results also strongly support this premise. In Fig. \ref{fig:class_dependent_coteaching_plastic_backbone_detials} and Fig. \ref{fig:class_dependent_dividemix_plastic_backbone_details}, we compare the performance of self-supervised pretraining with both CT and DM across five datasets. Self-supervised pretraining improves performance by a relatively small margin compared to that with symmetrical label noise. The performance is limited by LNL rather than having a good pretrained feature. However, for DermNet, which has 23 classes grouped into three, so a large number of classes are dependent within the group, and symmetry is high. Therefore, we still see improvement offered by self-supervised pretraining. \textit{In a nutshell, self-supervised pretraining is more applicable and improves robustness against noisy labels more with symmetrical label noise than with class-dependent label noise.}

\subsection{Self-supervised Pretraining on ImageNet weights}
\label{Self-supervised Pretraining on ImageNet Pretrained}
\begin{figure}[h!]
    \includegraphics[width=0.9\linewidth]{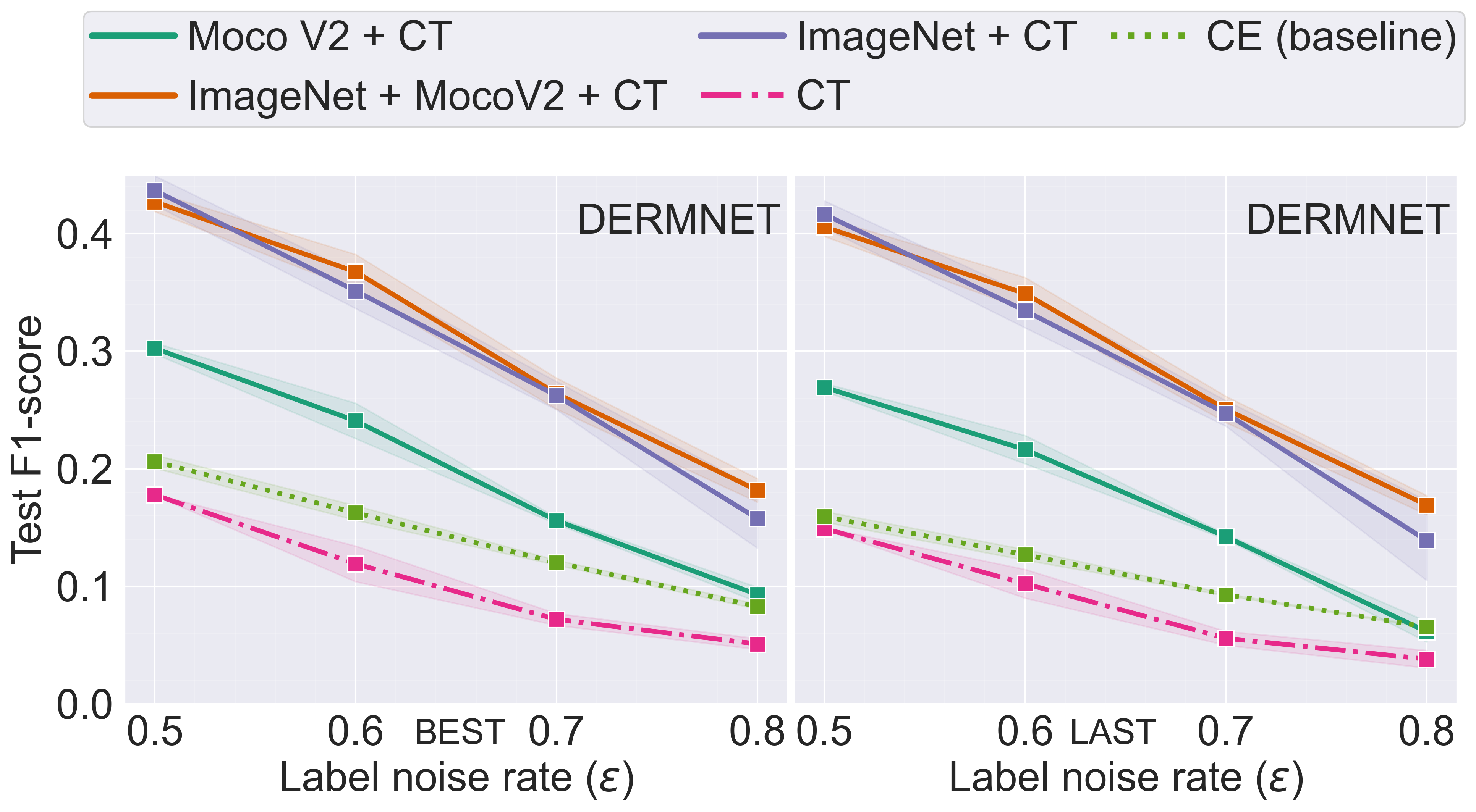}
    \caption{Comparing the test performance of Co-teaching (CT) with MoCo v2 pretrained from ImageNet weight (\textbf{plastic backbone}) against baselines at different \textbf{class-dependent} label noise rates in DermNet dataset. CE stands for standard cross-entropy. LAST and BEST show the best performance and average of the last five epochs, respectively in the test set.}
    \label{fig:coteaching_plastic_backbone_imagenet_moco}
\end{figure}

For all the above experiments, we investigated which self-supervised pretraining approach yields greater robustness against noisy labels. It was important to pretrain the models from scratch for a fair study without inducing any bias from already pretrained weights on natural images. However, in real applications, we may want to leverage the strength of a pretrained model trained on a large out-of-domain dataset like ImageNet. Especially, if the domain gap between the out-of-domain and the experimental dataset is small, the self-supervised learning might benefit more from leveraging pretrained weight than training from scratch. Therefore, we pretrained MoCo v2 using ImageNet pretrained weights instead of training from scratch. In Fig. \ref{fig:coteaching_plastic_backbone_imagenet_moco}, we show the results of MoCo v2 + ImageNet pretraining with CT. We can clearly see that this strategy further improves the robustness of the self-supervised method against noisy labels.

\section{Discussion}

In our investigation, we examined five 2D medical datasets, encompassing various modalities such as X-ray, ultrasound, and RGB images. Due to differences in the number of classes and dataset sizes across these datasets, our initial analysis focused on accounting for these factors to fairly compare datasets and estimate their difficulty and robustness against noisy labels. Determining dataset difficulty is not straightforward, as demonstrated in Section \ref{dataset difficulty}, and can vary depending on the evaluation criteria. Although all analyses identified DermNet as the most challenging dataset, the ranking of other datasets was not consistent across all analyses. Interestingly, DermNet exhibited the highest robustness against noisy labels among all the datasets. While the performance of other datasets dropped abruptly after a certain threshold, DermNet showed a linear drop even when the same number of classes were grouped into three.

Additionally, we observed that contrastive learning-based pretraining, which demonstrated the best performance against noisy labels, improved even further with extended training. However, for other methods, overtraining had a detrimental effect on performance. In generative approaches, overtraining led to model collapse and posterior collapse, resulting in poor feature representation. Another noteworthy observation was that pretraining provided more benefits in cases of symmetrical label noise, whereas only marginal benefits were seen in class-dependent label noise. It is crucial to note that pretraining no longer offers an advantage beyond the label noise rate flipping threshold. For instance, in COVID-DU-Ex at $\epsilon = {0.7,0.8}$, it performed worse than random labels, indicating that self-supervised pretraining didn't provide significant assistance in this scenario. Finally, as discussed in Section \ref{Self-supervised Pretraining on ImageNet Pretrained}, we demonstrated that the most effective strategy is to apply self-supervised pretraining on top of a model pretrained on a large out-of-domain dataset like ImageNet if the domain gap between out-od-domain dataset and experimental dataset is minimum.

However, this work has some limitations. State-of-the-art LNL methods like Coteaching and Dividemix were primarily designed for fairly balanced datasets. Since none of the investigated datasets are severely class-imbalanced or long-tailed, CT and DM are suitable choices. Nevertheless, these methods may not perform well on severely imbalanced datasets, which could be an interesting direction for future research.

Additionally, our analysis was limited to a CNN-based backbone (ResNet18) to maintain experiment manageability and ensure fair comparisons across all methods. With transformers gaining popularity as backbones in recent years, it is likely that their behavior against label noise differs from that of CNNs. Therefore, further research should investigate the performance of downstream classification tasks using a transformer-based architecture—a topic we plan to explore in future works.

\section{Conclusion}

In conclusion, our investigation delved into various self-supervised pretraining approaches in conjunction with supervised LNL methods to assess their potential for enhancing robustness in medical image classification. We examined five diverse datasets that encompassed variations in training samples, class numbers, and imaging modalities. Our study revealed that contrastive learning emerged as the most effective among all self-supervised pretraining methods in improving robustness against noisy labels when combined with existing supervised techniques. Furthermore, we conducted an in-depth analysis of the impact of noisy labels on medical image classification, taking into account factors such as dataset complexity, class count, and dataset size. This paper offers a comprehensive study and valuable insights that can be applied to adapt self-supervised pretraining in diverse settings and address label noise challenges across different characteristic datasets.

\section*{Declaration of competing interest} The authors declare that they have no known competing financial interests or personal relationships that could have appeared to
influence the work reported in this paper.

\section*{Data Availability} We used publicly available datasets. We will make the code public during publication. \\

\noindent\textbf{Acknowledgements.} The research reported in this publication was supported by the National Institute of General Medical Sciences Award No. R35GM128877 of the National Institutes of Health, the Office of Advanced Cyber Infrastructure Award No. 1808530 of the National Science Foundation, and the Division Of Chemistry, Bioengineering, Environmental, and Transport Systems Award No. 2245152 of the National Science Foundation. We would like to thank the Research Computing team at the Rochester Institute of Technology \citep{RITRC} for providing computing resources for this research. 
\bibliographystyle{model2-names.bst}\biboptions{authoryear}
\bibliography{refs}

\end{document}